%% file: ClippingK1000.tex
\documentclass[fleqn,usenatbib]{mnras}

\usepackage{newtxtext,newtxmath}
\usepackage[T1]{fontenc}

\DeclareRobustCommand{\VAN}[3]{#2}
\let\VANthebibliography\thebibliography
\def\thebibliography{\DeclareRobustCommand{\VAN}[3]{##3}\VANthebibliography}

\usepackage{graphicx}	% Including figure files
\usepackage{amsmath}	% Advanced maths commands
\usepackage{gensymb}

\newcommand{\LCDM}{$\Lambda$CDM}

\title[Clipping KiDS-1000]{KiDS-1000: Improved constraints on cosmology, intrinsic alignments and baryonic feedback from clipped cosmic shear}

\author[B. Giblin et al.]{
Benjamin Giblin,$^{1}$\thanks{E-mail: bengib@roe.ac.uk}
Joachim Harnois-D{\'e}raps$^{2}$, 
Louise Paquereau$^{3}$, 
and Catherine Heymans$^{1,4}$
\\
$^{1}$Institute for Astronomy, University of Edinburgh, Royal Observatory, Blackford Hill, Edinburgh, EH9 3HJ, UK \\
$^{2}$School of Mathematics, Statistics and Physics, Newcastle University, Herschel Building, NE1 7RU, Newcastle-upon-Tyne, UK\\
$^{3}$Department of Physics and Astronomy, Astronomy and Plasma Physics (AoP) Division, Chalmers University of Technology, Gothenburg, Sweden \\
$^{4}$Ruhr University Bochum, Faculty of Physics and Astronomy, Astronomical Institute (AIRUB), German Centre for Cosmological Lensing, 44780 Bochum, Germany \\
}

\date{Accepted XXX. Received YYY; in original form ZZZ}

\pubyear{\the\year{}}

\begin{document}
\label{firstpage}
\pagerange{\pageref{firstpage}--\pageref{lastpage}}
\maketitle

\begin{abstract}
We present improved cosmological constraints from the fourth data release of the Kilo-Degree Survey (``KiDS-1000") using \textit{clipped} shear correlation functions. Clipping filters the projected density field inferred from weak lensing data for the highest-density regions, allowing for two-point functions to extract additional cosmological information. We model, for the first time, the impact of systematics on clipped lensing statistics, including intrinsic alignments, baryonic feedback, photometric redshift uncertainties, and source-lens clustering. We train Gaussian process emulators on dark-matter-only and hydrodynamical $N$-body simulations to predict the cosmological and systematic dependence of both the clipped and conventional, ``unclipped" shear correlation functions. We find that the combination of the clipped and unclipped probes improves the constraints on the free parameters of the $w$CDM model relative to the conventional approach, with a 16\% tightening of the $S_8$ uncertainty and 24\% for $w_0$. Our constraints, $\Omega_{\rm m} = 0.263^{+0.035}_{-0.038}$, $S_8=0.724^{+0.027}_{-0.027}$, and $w_0 = -1.24^{+0.26}_{-0.28}$, are consistent with the $\Lambda$CDM model and with the cosmic shear analysis of \cite{asgari/etal:2021} via a completely independent simulation- and emulator-based forward-modelling approach. We also find the complimentary information in the clipped statistic provides an upper limit on the baryon feedback strength, and improves the constraints on intrinsic alignments by 27\%.
\end{abstract}


%
%________________________________________________________________

\input{Section_1_Introduction}

\input{Section_2_Data_Sims}

\input{Section_3_Method}

\input{Section_4_Results}
\input{Section_5_Conclusions}

\section*{Acknowledgements}
We thank Mike Jarvis for the excellent {\sc TreeCorr} software package used in computing clipped and unclipped shear correlation functions in this analysis. We are also very grateful to Eric Tittley for maintaining the {\sc cuillin} HPC system at the Institute for Astronomy, which made this work possible, and to Tiago Castro, Klaus Dolag and Nicolas Martinet for their contributions to the production of the KiDS-1000 {\sc magneticum} simulations.

BG acknowledges support from the UKRI Stephen Hawking Fellowship (grant reference EP/Y017137/1). JHD acknowledges support from an STFC Ernest Rutherford Fellowship (project reference ST/S004858/1) for the earlier part of this work. CH acknowledges support from the Max Planck Society and the Alexander von Humboldt Foundation in the framework of the Max Planck-Humboldt Research Award endowed by the Federal Ministry of Education and Research, and the UK Science and Technology Facilities Council (STFC) under grant ST/V000594/1.

The software packages which, respectively, applied clipping to the KiDS-1000 data and simulations, and trained the GP emulators before running the parameter inference chains, are open-source and available at: \href{https://github.com/benjamingiblin/ClippingPipeline}{https://github.com/benjamingiblin/ClippingPipeline} and \href{https://github.com/benjamingiblin/Calc_Lhd_Tool}{https://github.com/benjamingiblin/Calc\textunderscore Lhd\textunderscore Tool}. The {\sc SLICS} and {\sc cosmoSLICS} are available at \href{https://slics.roe.ac.uk/}{https://slics.roe.ac.uk/}. The KiDS-1000 data are available at \href{kids.strw.leidenuniv.nl/DR4/lensing.php}{kids.strw.leidenuniv.nl/DR4/lensing.php}. 
  
The results in this paper are based on observations made with ESO Telescopes at the La Silla Paranal Observatory under programme IDs 177.A-3016, 177.A-3017, 177.A-3018 and 179.A-2004, and on data products produced by the KiDS consortium. The KiDS production team acknowledges support from: Deutsche Forschungsgemeinschaft, ERC, NOVA and NWO-M grants; Target; the University of Padova, and the University Federico II (Naples). Contributions to the data processing for VIKING were made by the VISTA Data Flow System at CASU, Cambridge and WFAU, Edinburgh. The {\sc magneticum} simulations were carried out at the Leibniz Supercomputer Center (LRZ) under the 
project pr83li. \\

\bibliographystyle{mnras} % style aa.bst
\bibliography{references} % your references 

\appendix
\input{Section_Appendix}

%-------------------------------------------------------------------
\end{document}

%% file: Section_1_Introduction.tex
\section{Introduction}
\label{sec:intro}

The Standard Model of Cosmology, {\LCDM}, continues to be the most successful description of the structure of our Universe and its evolution from early to late times. Despite this fact, the natures of dark matter and dark energy - the two most abundant sources of energy density in the {\LCDM}  model - remain elusive, and debates over the values of the cosmological parameters of the model persist.  Constraints on the Hubble parameter, $H_0$, informed by the redshift-distance relation of low-redshift standard candles \citep{riess/etal:2021, breuval/etal:2024} are consistently reported at 4--6$\sigma$ higher than those derived from the high-redshift temperature and polarisation fluctuations in the CMB \citep{planck/etal:2018}. As for the value of $S_8=\sigma_8 \sqrt{ \Omega_{\rm m} / 0.3}$, where $\sigma_8$ quantifies the variance of matter density in spheres of 8 Mpc/$h$ and $\Omega_{\rm m}$ is the matter energy density parameter, the constraints from Stage-III weak lensing surveys range from consistency with CMB-derived estimates \citep{wright/etal:2025} to mild/moderate tensions at the level of 2--3$\sigma$ \citep{dalal/etal:2023, sugiyama/etal:2025, des/etal:2026}. The questions raised by these cosmological tensions concerning the validity of {\LCDM} in simultaneously modelling high- and low-redshift observations, are further compounded by the statistical preference for a time-varying dark energy found in the latest clustering measurements of the Dark Energy Spectroscopic Instrument \citep{desi/etal:2025f,desi/etal:2025g}.

In this era of precision cosmology, where ever-increasing survey data volumes have driven statistical uncertainties below those from systematics in many cases, the necessity for robust control of systematics is paramount. In the case of weak lensing, in addition to the detector-based systematics affecting galaxy shape measurement, the cosmological lensing signal (aka `cosmic shear') is contaminated by \textit{intrinsic alignments} of galaxies \citep[IA; see for example,][]{heavens/etal:2000,hirata/etal:2004,heymans/etal:2013}, baryonic feedback \citep{semboloni/etal:2011b}, photometric redshift uncertainties \citep{zhang/etal:2010}, and potentially, coupling of these effects \citep{bridle/king:2007,leonard/etal:2024}, to name but a few \citep[see ][for a comprehensive overview]{mandelbaum:2018}. 

The tantalising hints of cosmological tensions from lensing surveys further call for improving the precision of constraints derived from this probe. Cosmic shear practitioners conventionally employ two-point summary statistics, such as shear correlation functions, lensing power spectra, or COSEBIs modes \citep[see, for example,][]{asgari/etal:2021}, to derive cosmological constraints, on account of the relative maturity of efforts to model their sensitivity to the aforementioned systematics. However, two-point statistics extract information only up to second-order. Given that the late-time projected matter density distribution probed by lensing is highly non-Gaussian, and hence contains higher-order information, the cosmological constraints derived from two-point lensing probes are sub-optimal.

This fact has motivated considerable exploration of \textit{Higher Order Weak Lensing Statistics} (HOWLS), including density peak and void statistics \citep{martinet/etal:2021,davies/etal:2020,davies/etal:2021}, topological metrics like Minkowski functionals and persistent homology \citep{petri/etal:2015,heydenreich/etal:2021,grewal/etal:2022}, one-point convergence probability density functions \citep{boyle/etal:2021,thiele/etal:2020,giblin/etal:2023,castiblanco/etal:2024}, and many others \citep[see ][for summary]{euclid-howls/etal:2023}. Whilst these approaches have been shown to consistently outperform two-point functions in cosmological constraining power by tens of percent, the small scales targeted by HOWLS in the pursuit of gleaning additional cosmological information are the same scales on which the contamination from baryonic feedback is strongest. Consequently, the challenges in jointly modelling the dependence of HOWLS on cosmological and nuisance parameters has so far delayed HOWLS from replacing two-point probes as the lensing statistics of choice. 

This analysis contributes to the growing body of research aimed at moving beyond forecasts and establishing alternative lensing statistics as viable and robust means of extracting cosmological information whilst controlling for systematics with concurrent survey data \citep[see, for example,][]{gatti/etal:2021,harnois-deraps/etal:2024,marques/etal:2024,novaes/etal:2025}. In this pursuit, we derive cosmological constraints from the fourth data release of the Kilo-Degree Survey (``KiDS-1000") using  \textit{clipped} shear correlation functions. 

Clipping removes regions from the observed field above a given density threshold, targeting the high-end tails where the density distribution deviates most strongly from Gaussian. This process decorrelates Fourier modes, allowing for more efficient information extraction with two-point statistics and gains in precision when the clipped and original (``unclipped") statistics are modelled jointly. Application of clipping to galaxy clustering observations has led to improvements in the precision of galaxy bias, $\sigma_8$, and growth of structure constraints \citep{simpson/etal:2011,simpson/etal:2013,simpson/etal:2016,wilson:2016}. Clipping was also employed by \cite{lombriser/etal:2015} to derive stronger constraints on chameleon and Vainshtein screening mechanisms in modified gravity.

Analyses in which clipping has been applied to the projected lensing density fields (convergence) remain few. 
A proof-of-concept analysis by \cite{simpson/etal:2015} predicted a 3-fold improvement in the $\Omega_{\rm m}$--$\sigma_8$ figure-of-merit from measuring the power spectra of clipped noise-free lensing fields, whilst the first application of clipping to real lensing data (the third data release from the Kilo-Degree Survey, ``KiDS-450") in \citet[][`G18' hereafter]{giblin/etal:2018}  found a 17\% improvement in $S_8$, albeit without contending with systematics or using tomographic redshift binning. 
Whilst HOWLS generally aim for tighter constraints by targeting the small-scale non-linear regions of observations - those most polluted by baryonic feedback - the improvements from clipped lensing instead come from boosting the information extraction with two-point statistics from larger, linear scales, where there is less contamination from baryons.

This work builds on the foundation laid in G18, extending the modelling of clipped shear correlation functions to account for systematic effects including intrinsic alignments, baryonic feedback, photometric redshift uncertainties, masking, and source clustering. We improve the modelling of clipped statistics with more advanced numerical simulations \citep[the {\sc cosmoSLICS};][]{harnois-deraps/etal:2019} and introduce tomographic redshift binning within a clipped analysis for the first time. This work adopts an emulator-based approach, wherein all systematics and cosmological signal are forward modelled with $N$-body simulations. As such, our constraints from the traditional ``unclipped" shear correlation functions also serve as a validation of those derived via a fully independent pipeline and modelling framework in the main KiDS-1000 cosmic shear analysis \citep[][hereafter `A21']{asgari/etal:2021}. 

We introduce our simulations and the KiDS-1000 data in Sec.~\ref{sec:data}, outline our clipping methodology and tests for validating the accuracy of our modelling in Sec.~\ref{sec:method}, present results in Sec.~\ref{sec:results} and conclude in Sec.~\ref{sec:conc}.

%% file: Section_2_Data_Sims.tex
\section{Data and Simulations}
\label{sec:data}

\subsection{KiDS-1000}

\begin{figure*}
\begin{center}
\includegraphics[width=\textwidth]{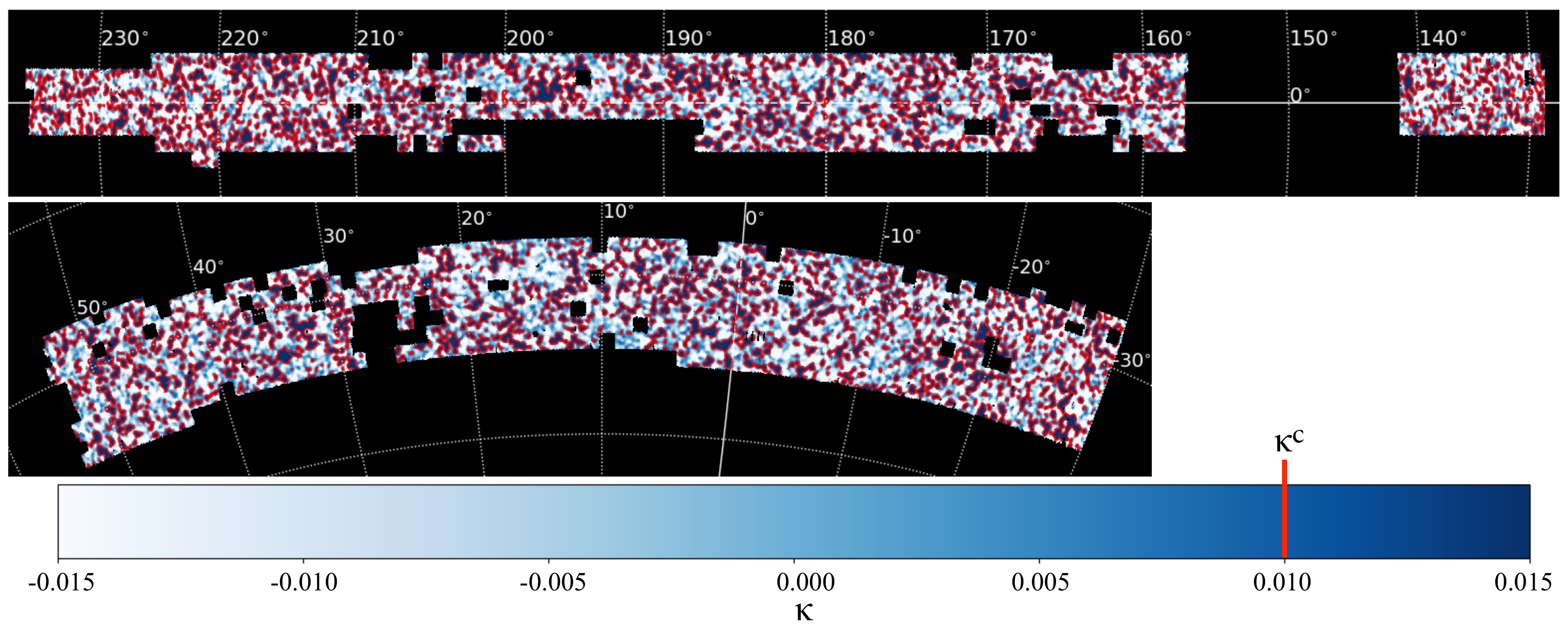}
\caption{Convergence, $\kappa$, maps for the northern (upper) and southern (lower) patches of KiDS-1000, measured from galaxies in the highest tomographic bin ($z_{\rm B} \in [0.9,1.2]$) with regions exceeding the clipping threshold ($\kappa^c = 0.010$) highlighted by the red contours (approximately 20\% of the observed area).}\label{fig:massmaps}
\end{center}
\end{figure*}

KiDS-1000, the publicly-available fourth data release from the Kilo-Degree Survey\footnote{\href{http://kids.strw.leidenuniv.nl/DR4/lensing.php}{kids.strw.leidenuniv.nl/DR4/lensing.php}}, contains ellipticity and photometric redshift (photo-$z$) estimates for 21 million galaxies. The footprint of the survey, shown in Figure \ref{fig:massmaps}, spans 1006 and 777 deg$^2$ of sky area before and after masking. The galaxy shape measurements were performed on the $r$-band imaging from the Survey Telescope on the VLT (the `VST') with a $5\sigma$ limiting magnitude of $r= 25.02 \pm 0.13$ and mean seeing of 0.7 arcsec. One of the main strengths of KiDS is the large number of bands used for photo-$z$ estimation, with four VST optical bands, $ugri$, and a further five near-infrared bands, $ZYJHK_{\rm s}$, provided by the overlapping matched-depth imaging of the VIKING (VISTA Kilo-Degree Infrared Galaxy) survey \citep{kuijken/etal:2019}. Calibration of the photo-$z$ distributions in KiDS-1000 \citep{hildebrandt/etal:2020,hildebrandt/etal:2021}, further benefited from multiple overlapping spectroscopic surveys including (but not limited to) the Baryon Oscillation Spectroscopic Survey \citep[BOSS;][]{dawson/etal:2013}, the Galaxy And Mass Assembly survey \citep[GAMA;][]{driver/etal:2011}, 2dFLenS  \citep{blake/etal:2016} and WiggleZ  \citep{drinkwater/etal:2010}. Hence, photometric redshift distributions were measured and calibrated for five tomographic redshift bins with edges, $[0.1,0.3,0.5,0.7,0.9,1.2]$. The robustness of the galaxy redshift and shear catalogues of KiDS-1000 to systematics including PSF mis-modelling and photo-$z$ errors was investigated in \cite{giblin/etal:2021} and found to meet all requirements. This enabled the KiDS-1000 cosmic shear analysis of \cite{asgari/etal:2021} and the combined lensing and galaxy clustering ($3\times 2$-point) analysis in \cite{heymans/etal:2020}. 

The fifth and final data release from KiDS - KiDS-Legacy \citep{wright/etal:2025} - benefits from a 33\% increase in area and improved redshift calibration. Legacy-tailored $N$-body simulations were, however, still under construction at the time this analysis was carried out; we therefore leave clipped lensing with KiDS-Legacy for future work.

\subsection{Simulations} \label{subsec:sims}

In this work we primarily rely on the {\sc cosmoSLICS} \citep{harnois-deraps/etal:2019}, and its sister suite, {\sc SLICS} \citep{harnois-deraps/etal:2018} - dark-matter-only $N$-body simulations which were specially designed to enable HOWLS analyses of weak lensing data from KiDS and the Dark Energy Survey \citep{harnois-deraps/etal:2024}, as well as the forthcoming Stage-IV lensing surveys \citep{euclid-howls/etal:2023}. {\sc cosmoSLICS} and {\sc SLICS} provide $\sim$47$\times$ larger simulation volume, twice the mass resolution, and >$200\times$ more particles than the \cite{dietrich/hartlap:2010} simulations used in G18 to model clipped lensing statistics. The {\sc SLICS} and {\sc cosmoSLICS} are also competitive with other mocks used to model HOWLS for Stage-III lensing surveys: {\sc CosmoGrid} \citep{kacprzak/etal:2023}, the \cite{marques/etal:2024} simulations, and {\sc Gower St} \citep{jeffrey/etal:2025}, offering more accurate reconstruction of non-linear scales than the former two suites and wider coverage of the cosmological parameter space than the latter two, generally at the cost of fewer cosmologies.

We use $N$-body simulations in this work in light of the absence of an accurate analytical prescription for clipped lensing statistics as a function of cosmological and nuisance parameters (a challenge which is faced generally by many HOWLS). We therefore train Gaussian process emulators on the following variations of the simulations to model the broad range of cosmological, astrophysical and survey-specific factors which shape the observed clipped shear correlation functions. For more detailed summaries of these mocks, we refer the interested reader to \cite{harnois-deraps/etal:2024}. 

 \begin{figure}
\begin{center}
\includegraphics[width=0.49\textwidth]{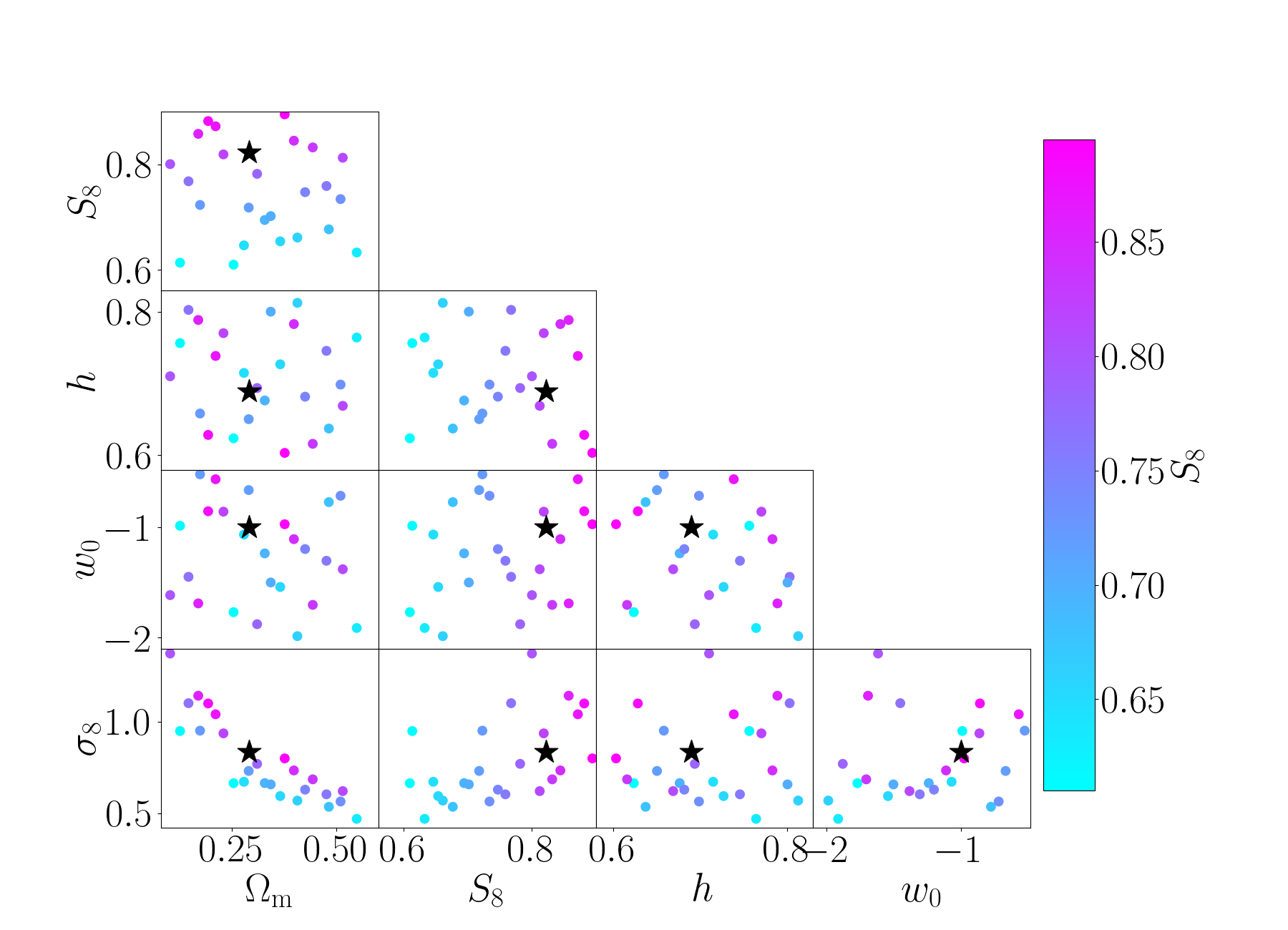}
\caption{The distribution of input parameters in the \textit{Cosmology Set} ({\sc cosmoSLICS}) colour-coded by $S_8 = \sigma_8 \sqrt{ \Omega_{\rm m}/0.3}$. The black star designates the fiducial cosmology at which the covariance matrix is estimated. } \label{fig:nodes}
\end{center}
\end{figure}  

(i) \textit{The Cosmology Set} are derived from the {\sc cosmoSLICS} $N$-body simulations and consist of 26 distinct $w$CDM cosmologies sampling, via a Latin hypercube, the 4-dimensional $\left( \Omega_{\rm m}, S_8, h, w_0 \right)$ parameter space, where $h=H_0 /  \left( 100 \, {\rm kms^{-1} \, Mpc^{-1}} \right)$ and $w_0$ is the present-day dark energy equation of state, as shown in Figure \ref{fig:nodes}. 
%The ranges of each sampled parameter are as follows: $\Omega_{\rm m} \in [0.10,0.55]$, $S_8 \in [0.60,0.90]$, $h \in [0.60, 0.82]$, $w_0 \in [-2.0, -0.5]$. 
For each cosmology, there are 50 realisations comprising 25 pairs with random seeds selected to suppress sample variance in the pairs' two-point functions, thereby optimising the accuracy of the average two-point function given the limited number of realisations. 

From each realisation in our cosmology training set, a simulated KiDS-1000 survey was constructed. This required the 777 deg$^2$ of observed area in KiDS-1000 (shown in Figure \ref{fig:massmaps}) to be divided into 18 tiles, each spanning the $10 \times 10$ deg$^2$ projected lightcone of {\sc cosmoSLICS}, and containing on average $\simeq 43$ deg$^2$ of observed sky after masking. Each {\sc cosmoSLICS} realisation was then used to create an 18-tile simulated KiDS-1000 survey by placing galaxies at the angular positions and photometric redshifts of those found in the actual data tiles, before ray tracing through the simulated volume to create lensing maps. Given that this process creates a mock survey in which the same large-scale structure is repeated across the 18 tiles (a feature which is not present in the real data), the tiles were then shuffled across the 50 cosmological realisations in order to create 50 mock KiDS-1000 surveys with no realisation appearing more than once across the 18 tiles. The absence of correlations across tiles in the simulations necessitates that we compute per-tile clipped and unclipped shear correlations functions which we then average (weighted by galaxy number counts) across the 18 tiles; for consistency we compute the KiDS-1000 measurements in the same way. 

The replication of the KiDS-1000 observed footprint in the mocks means our model predictions are affected by masking in the same way as the KiDS-1000 data. Whilst masking has negligible impact on the traditional \textit{unclipped} shear correlation functions, it was demonstrated in G18 to bias clipped statistics by tens of percent. An empirical masking correction was developed in G18 and found to reduce the bias in the clipped $\xi_+$ to $<5\%$ but the clipped $\xi_-$ continued to be affected at the level of $\sim 10\%$ and hence, was not included in the parameter inference. By infusing the KiDS-1000 footprint into the mocks, in addition to the survey's redshift distributions and galaxy number densities, the mask bias is forward modelled in the parameter inference, facilitating the use of both components of the clipped shear correlation functions for the first time.

The intrinsic galaxy shape noise level of KiDS-1000 are also replicated in the mocks by infusing the simulated shears with intrinsic ellipticities obtained by applying random rotations to the corresponding galaxies' ellipticities in KiDS-1000, thereby erasing the cosmological signal in the data. The redshift-bin-dependent multiplicative, $m$-, and additive, $c$-, shear corrections estimated for KiDS-1000 \citep[see Table 1 of ][]{giblin/etal:2021} are also applied to our simulated galaxies, with the caveat that a \textit{per-object} $m$-correction, derived from the object's apparent size, magnitude and observed galaxy shape using the relation in \cite{kannawadi/etal:2019} is applied in this work, whereas a single $m$-correction per tomographic bin was adopted in A21. The per-object correction aims to forward model the local variations in noise across the KiDS footprint that can correlate with the galaxy weights (these are also lifted from the KiDS-1000 data and applied to the mocks).

(ii) \textit{The Covariance Set} of simulations are derived from the {\sc SLICS}, a suite which matches the {\sc cosmoSLICS} in all specifications save for the fact all realisations are completely independent (i.e. have no paired initial conditions) and are at a fixed cosmology matching one of the nodes in {\sc cosmoSLICS}, shown by the black star in Fig.~\ref{fig:nodes} at the approximate centre of the parameter space: $\Omega_{\rm m}=0.2905, S_8=0.8231, h=0.6898, w_0=-1.0$ (hereafter referred to as the `fiducal' cosmology). Specifically this suite consists of 124 independent simulated KiDS-1000 surveys constructed in the same manner as the Cosmology Set, providing an excellent basis to calculate the covariance matrices of the clipped and unclipped statistics. ~Our approach to parameter inference hence adopts the common ansatz of a cosmology-independent covariance. Biases arising from this assumption have been shown to be small relative to the statistical uncertainty of Stage-III-like surveys such as KiDS \citep{eifler/etal:2009}.

In order to further augment the number of realisations in the Covariance Set, leading to a more stable covariance inversion in the likelihood, we produce 10 galaxy shape noise configurations for every cosmological realisation, yielding 1,240 mock KiDS-1000 surveys. 

(iii) \textit{The Intrinsic Alignments (IA) Set}, first presented in \cite{harnois-deraps/etal:2022}, were produced by infusing the galaxies in every realisation of the fiducial cosmology simulations of {\sc cosmoSLICS} with the so-called `non-linear linear alignment' (NLA) model \citep{bridle/king:2007}. This is implemented by generating per-object IA contributions to the galaxy ellipticity components which are linearly coupled with the projected tidal tensor at each galaxy position, with the coupling strength set by the amplitude parameter, $A_{\rm IA}$. The positions of galaxies are also sampled so as to trace the dark matter density, $\delta$, with a linear galaxy bias of $b_{\rm gal}=1.0$. This means the simulated IA signal is consistent with the `$\delta$-NLA' model developed in \cite{blazek/etal:2019} to account for the biased sampling of the IA field at galaxy positions. This assumes no luminosity or redshift dependence of the IAs and therefore lacks the full complexity of the Tidal Alignment and Torquing (TATT) model \citep{blazek/etal:2019}. Neglecting the additional parameters of TATT, however, is justified given that they are unconstrained by the statistical power of Stage-III weak lensing surveys \citep{secco/etal:2021}.

Our IA Set thus consists of 50 pseudo-independent realisations  and features the KiDS-1000 photometric redshift distributions and galaxy shape noise levels. The KiDS footprint, however, is not replicated in these mocks with each realisation spanning the full $10 \times 10$ deg$^2$ lightcone. The use of the IA Set in this work, therefore, invokes the assumption that the IA and cosmological contributions to our statistics can be modelled independently, and that the impact of masking on the IA-bias to the clipped and unclipped correlation functions is negligible. Whilst this assumption is almost unavoidable, given the computational expense of producing simulations which jointly model cosmological and nuisance parameters, it is also very likely to be valid given that we use the IA Set only to compute the \textit{bias} to our predictions as the \textit{difference} between $A_{\rm IA} \neq 0$ and $A_{\rm IA}=0$ mock measurements. This isolates the impact of changing $A_{\rm IA}$ on the IA-bias whilst damping the contributions from the values of the cosmological parameters and the specifics of the masking.

(iv) \textit{The Photo-$z$ Set} are used to model the biases to our correlation functions from potential inaccuracies in the mean of the per-bin photometric redshift distributions, $n(z)$. These were produced by subjecting the simulation $n(z)$'s to shifts, $\Delta z$, sampled from Gaussian distributions with means and widths given by the KiDS-1000 redshift distribution bias and errors estimated in \cite{hildebrandt/etal:2021} and given in Table 1 of \cite{giblin/etal:2021}. The impact of these shifts on the $n(z)$ can be seen in Fig. 1 of \cite{harnois-deraps/etal:2024}. In total, four shifts of the $n(z)$ were used to construct the Photo-$z$ Set, given by $\boldsymbol{\Delta z} = \left( -0.898, -0.229, 0.262, 0.8254 \right) \boldsymbol{\sigma_z}$, where $\boldsymbol{\sigma_z} = \left( 0.0096, \, 0.0114, \, 0.0116, \, 0.0084, \, 0.0097 \right)$ is the uncertainty on the bias to the mean of the redshift distribution for each of the five KiDS-1000 redshift bins.

10 of the pseudo-independent realisations from {\sc cosmoSLICS}  with the fiducial cosmology were infused with the $n(z)$ shifts corresponding to each $\Delta z$ value. Like the Cosmology and Covariance Sets, the KiDS-1000 footprint and galaxy positions are replicated in these mocks via 18 tiles, in addition to the intrinsic shape noise levels. As with the IA Set, these mocks are used only to infuse our predictions with the biases computed as the difference between measurements with ($\Delta z \neq 0$) and without ($\Delta z = 0$) shifts to the $n(z)$. 

(v) \textit{The Baryonic Feedback (BF) Set} are used to model biases to the clipped and unclipped shear correlation functions arising from the energetic output of baryonic sources and its impact on large-scale structure. This is modelled with the {\sc magneticum} simulations which use the smoothed particle hydrodynamics (SPH) code {\sc P-Gadget} \citep{springel:2005} to modulate the matter density field to account for AGN and supernovae feedback, radiative cooling and star formation \citep{castro/etal:2021}. The suppression of the matter power spectrum due to these processes as modelled by {\sc magneticum} was measured by \cite{martinet/etal:2021b} to be $\sim$15\% suppression on scales $k=10 \, h/{\rm Mpc}$ at $z=0$; similar in magnitude to the \citet{mccarthy/etal:2017} version of {\sc BAHAMAS} \citep[ran with the WMAP cosmology,][and with the temperature jump from AGN heating set to $T_{\rm AGN}=10^{7.8}{\rm K}$]{hinshaw/etal:2013}.

Our model for BF relies on 10 pseudo-independent $10 \times 10$ deg$^2$ lightcones featuring the full hydrodynamical physics of {\sc magneticum} and another 10 dark-matter-only mocks evolved from the same initial conditions. These lightcones were constructed with the {\sc slicer}\footnote{\url{https://github.com/TiagoBsCastro/SLICER}} 
software and are tailored to the KiDS-1000 specifications in the same way as the Cosmology, Covariance and Photo-z Sets. We use the BF-contaminated and BF-free simulations to compute BF-biases to our clipped and unclipped statistics analogously to the handling of the other aforementioned systematics. Given that there is a wide range of uncertainty on the impact of BF on the matter power spectrum \citep{chisari/etal:2018}, we follow \citet{harnois-deraps/etal:2024} and within the parameter inference linearly scale the amplitude of the BF-bias with a free nuisance parameter, $b_{\rm bary} \in [0.0, 2.0]$, allowing for a range of biases from zero to twice the BF levels of {\sc magneticum}.

%% file: Section_3_Method.tex
\section{Methodology}
\label{sec:method}

\subsection{Clipping} 

Our clipping methodology closely follows that which was developed in G18; hence, we briefly summarise the approach here, which is applied identically to the simulations as to the data, and we refer the reader to this publication for further details.

As explained in Sec.~\ref{sec:data}, our simulations are infused with the KiDS-1000 footprint via 18 10$\times$10 deg$^2$ tiles. Average ellipticity maps are constructed by projecting the galaxies (both simulated and observed) onto grids of pixels with resolution 2.3 arcmin per pixel; this scale corresponds to an average galaxy density of 3-10 pxl$^{-1}$ depending on the tomographic bin \citep{giblin/etal:2021}. Since each tile subtends a small sky area, we use the flat-sky \citet[][KS]{kaiser/squires:1993} mass mapping approach to produce convergence, $\kappa$, maps. We first smooth the ellipticity maps with a Gaussian filter of width $\sigma_{\rm s}=6.6$ arcmin and pad with a border of zeros to suppress the impacts of noise and edge effects on the derived $\kappa$ values \citep{vanwaerbeke/etal:2013}. We then perform the KS-reconstruction and reapply the mask to the resultant $\kappa$ map to reset the values of unobserved pixels.

The convergence in each pixel of the map, $\boldsymbol{\theta}$, is then subjected to clipping if above a given threshold, $\kappa^c$:

\begin{equation}
\kappa^{\rm{clip}} (\boldsymbol{\theta}) =  
\begin{cases}
\kappa^c, & \text{if } \kappa (\boldsymbol{\theta}) \geq \kappa^c  \\ \kappa (\boldsymbol{\theta}), & \text{otherwise}  
\end{cases}\,.
\end{equation}

G18 performed a detailed analysis of the impact of different clipping thresholds and Gaussian smoothing scales on both the clipped convergence maps and the derived shear correlation functions. Whilst a degeneracy was observed between these two free parameters, $\kappa^c=0.010$ and $\sigma_{\rm s}=6.6$ arcmin were found to strike a good balance by isolating high-density regions of the field where non-Gaussianity begins to emerge without clipping the underlying linear signal. This analysis was performed for the previous KiDS data release, KiDS-450, and in a single broad redshift bin, $z_{\rm B} \in [0.5, 0.9]$. Hence we repeated one of the tests from G18 to verify that these values continue to target the non-Gaussian regions in all of the tomographic bins of our KiDS-1000-like simulations. This consisted of checking that the clipped correlation functions exhibit a loss of power on small scales (those clipping is designed to impact) but recover the large-scale behaviour of its unclipped counterpart. This is not a strict requirement for all models in our Cosmology Set of simulations; in fact, a failure to recover the large-scale power of the unclipped correlation function is a useful indicator of extremal $S_8$ values and highlights the power of clipping for discriminating between different cosmologies. Nevertheless, we find $\kappa^c=0.010$ and $\sigma_{\rm s}=6.6$ arcmin do indeed pass this test across all redshift bins for the majority of cosmologies, and therefore adopt these values going forward. 

 Figure \ref{fig:massmaps} shows the convergence maps obtained from the northern (upper) and southern (lower) KiDS-1000 footprints using galaxies in the highest signal-to-noise redshift bin, $z_B \in [0.9,1.2]$, with the areas impacted by clipping ($\kappa > 0.010$) highlighted by the red contours. The fraction of sky which is clipped, $\sim$20\%, is very similar to that which was observed for KiDS-450 in G18, further supporting the continued use of the chosen clipping threshold and smoothing scale. 

Clipping essentially splits a $\kappa$ map into two components: the clipped map, $\kappa^{\rm clip}$, containing the ``lowlands" of the field (all features below the adopted threshold, $\kappa^c$, with planes of $\kappa=\kappa^{\rm c}$ where the peaks were previously), and a residual map, $\Delta \kappa$, containing the ``highlands" ($\kappa-\kappa^c$ values at the peaks surrounded by a sea of $\kappa=0$ values):
\begin{equation}
\kappa(\boldsymbol{\theta}) = \kappa^{\rm clip}(\boldsymbol{\theta}) + \Delta \kappa(\boldsymbol{\theta}) .
\end{equation}
The $\Delta \kappa$ map is then subjected to the inverse-KS process to recover two ``residual ellipticity" maps, $\Delta \epsilon(\boldsymbol{\theta}) = \Delta \epsilon_1 (\boldsymbol{\theta}) + i \Delta\epsilon_2 (\boldsymbol{\theta})$, which are near-zero across the majority of the field, but at the positions of convergence peaks contain the residual ellipticities associated with these overdense regions. The mask is then applied to ensure that all ``lowland" pixels (those not subjected to clipping) have $\Delta \epsilon(\boldsymbol{\theta})=0$ before performing a two-dimensional linear interpolation to obtain the residual ellipticities at the positions of galaxies in KiDS-1000 (and therefore also, our simulations), $\boldsymbol{\theta_g}$. Clipped ellipticities are then obtained from the observed unclipped ellipticities:
\begin{equation}
{\epsilon}^{\rm{clip}}(\boldsymbol{\theta}_{\rm{g}}) = \epsilon^{\rm{obs}}(\boldsymbol{\theta}_{\rm{g}}) - \Delta \epsilon(\boldsymbol{\theta}_{\rm{g}}).
\end{equation} 
The application of this mask to the $\Delta \epsilon(\boldsymbol{\theta})$ ensures that only the ellipticities for galaxies coinciding with convergence peaks are affected by clipping. 

Finally, we estimate the tomographic clipped and unclipped shear correlation functions in nine logarithmically-spaced angular separation bins, $\theta$, between 0.5 and 300 arcmin (the scales used in previous KiDS cosmic shear analyses, e.g., A21), as:
\begin{equation} \label{eqn:xi+-}
\widehat{\xi_{\pm}^{ij}}(\theta) = \frac{\sum_{\rm ab} w_{\rm a}^i w_{\rm b}^j \left[ \epsilon_{\rm t}^i (\boldsymbol{\theta}_{\rm g,a}) \epsilon_{\rm t}^j (\boldsymbol{\theta}_{\rm g,b}) \, \pm \, \epsilon_\times^i (\boldsymbol{\theta}_{\rm g,a}) \epsilon_\times^j (\boldsymbol{\theta}_{\rm g,b}) \right] }{ \sum_{\rm ab} w_{\rm a}^i w_{\rm b}^j } \,.
\end{equation}
Here the $\, \widehat{\,} \,$ notation indicates that the estimation is made in the presence of noise (galaxy ellipticities as opposed to shears), and $\epsilon_{\rm{t},\times}^i$ represent either the clipped or unclipped ellipticities in redshift bin $i$ measured tangentially (subscript `${\rm t}$') or at 45$\degree$ (subscript `$\times$') to the vector connecting galaxies at positions $\boldsymbol{\theta}_{\rm{g},\rm{b}}$ and $\boldsymbol{\theta}_{\rm{g},\rm{b}}$. The bin-dependent galaxy weights, $w_{\rm a,b}^i$, are from the KiDS galaxy shape measurement algorithm \textit{lens}fit \citep{miller/etal:2013,fenechconti/etal:2017}.

Auto- ($i=j$) and cross- ($i \neq j$) correlations are measured for the five KiDS-1000 redshift bins for unclipped and clipped ellipticities separately, generating a data vector of length 15 (redshift bin combinations) $\times$ 9 ($\theta$ bins) $\times$ 2 ($+$ \& $-$) $\times$ 2 (clipped \& unclipped) $=540$. One could in principle also measure cross correlations between the clipped and unclipped galaxy ellipticities, but this would dramatically increase the size of the data vector further and possibly render inverse-covariance estimation unstable for the sake of what is likely to be limited additional cosmological information.

As discussed further in Sections \ref{subsec:emu_train} and \ref{subsec:param_infer}, we use emulators trained on the simulation clipped and unclipped correlation functions to provide model predictions within our parameter inference framework. This is only strictly necessary for the clipped measurement, $\xi_\pm^{{\rm c}; ij}$, given that a fast and accurate framework for predicting the unclipped, $\xi_\pm^{{\rm uc};ij}$, is available and given by

\begin{equation} \label{eqn:xi+-theory}
\xi_\pm^{{\rm uc};ij}(\theta) = \frac{1}{2\pi}\int \text{d}\ell \, \ell \,P^{ij}_\kappa(\ell) \, J_{0,4}(\ell \theta) \, , 
\end{equation}

\noindent where the zeroth and fourth order Bessel functions, $J_{0,4}(\ell \theta)$, are used for $\xi_+^{\rm{uc}}$ and $\xi_-^{\rm{uc}}$ respectively. The convergence power spectrum for the $i$-$j$ redshift bin combination, $P^{ij}_\kappa(\ell)$, is given by

\begin{equation} \label{eqn:Pkappa}
P^{ij}_\kappa(\ell) = \int_0^{\chi_{\rm H}} \text{d} \chi \, \frac{q^i(\chi) q^j(\chi)}{f_{\rm{K}}(\chi)^2} \, P_\delta \left( k=\frac{[\ell+1/2]}{f_{\rm{K}}(\chi)},\chi \right),
\end{equation}

\noindent where $P_\delta(k,\chi)$ is the matter power spectrum, $\chi_{\rm{H}}$ is the comoving radial distance to the horizon and $k$ is the Fourier conjugate of $\chi$. The lensing efficiency, $q(\chi)$, is defined as

\begin{equation} \label{eqn:lensing_efficiency}
q^i(\chi) = \frac{3 H_0^2 \Omega_{\rm m}}{2c^2} \frac{f_{\rm{K}}(\chi)}{a(\chi)}\int_\chi^{\chi_{\rm H}}\, \text{d} \chi^\prime\ n^i(\chi^\prime) 
\frac{f_{\rm{K}}(\chi^\prime-\chi)}{f_{\rm{K}}(\chi^\prime)}\,,
\end{equation}
\noindent where $a$ is the scale factor, $n^i(\chi)$ is the redshift distribution of galaxies in the $i$'th bin and $c$ is the speed of light.

It would be possible, therefore, to follow the KiDS-1000 cosmic shear analysis (A21) and use this theoretical prescription, coupled with the non-linear matter power spectra predictions from {\sc hmcode} \citep{mead/etal:2016,mead/etal:2021} to model the unclipped shear correlation functions. Nevertheless, we implement the simulations also for the unclipped modelling in this analysis for three reasons: 1) By forward modelling the entire KiDS-1000 weak lensing survey with $N$-body simulations, this work serves as an independent validation of the {\sc hmcode}-based KiDS-1000 cosmic shear results of A21; 2) The cosmological parameter range spanned by {\sc cosmoSLICS} is wider than the range in the simulations used for calibrating {\sc hmcode} \citep{mead/etal:2021} and hence provides more reliable predictions for extremes in $\Omega_{\rm m}, S_8$ and $w_0$, and; 3) using {\sc cosmoSLICS} for both the clipped and unclipped $\xi_\pm$ facilitates a cleaner measure of the gain in constraining power facilitated by clipping, which is one of the key goals of this analysis. 

 \begin{figure*}
\begin{center}
\includegraphics[width=0.95\textwidth]{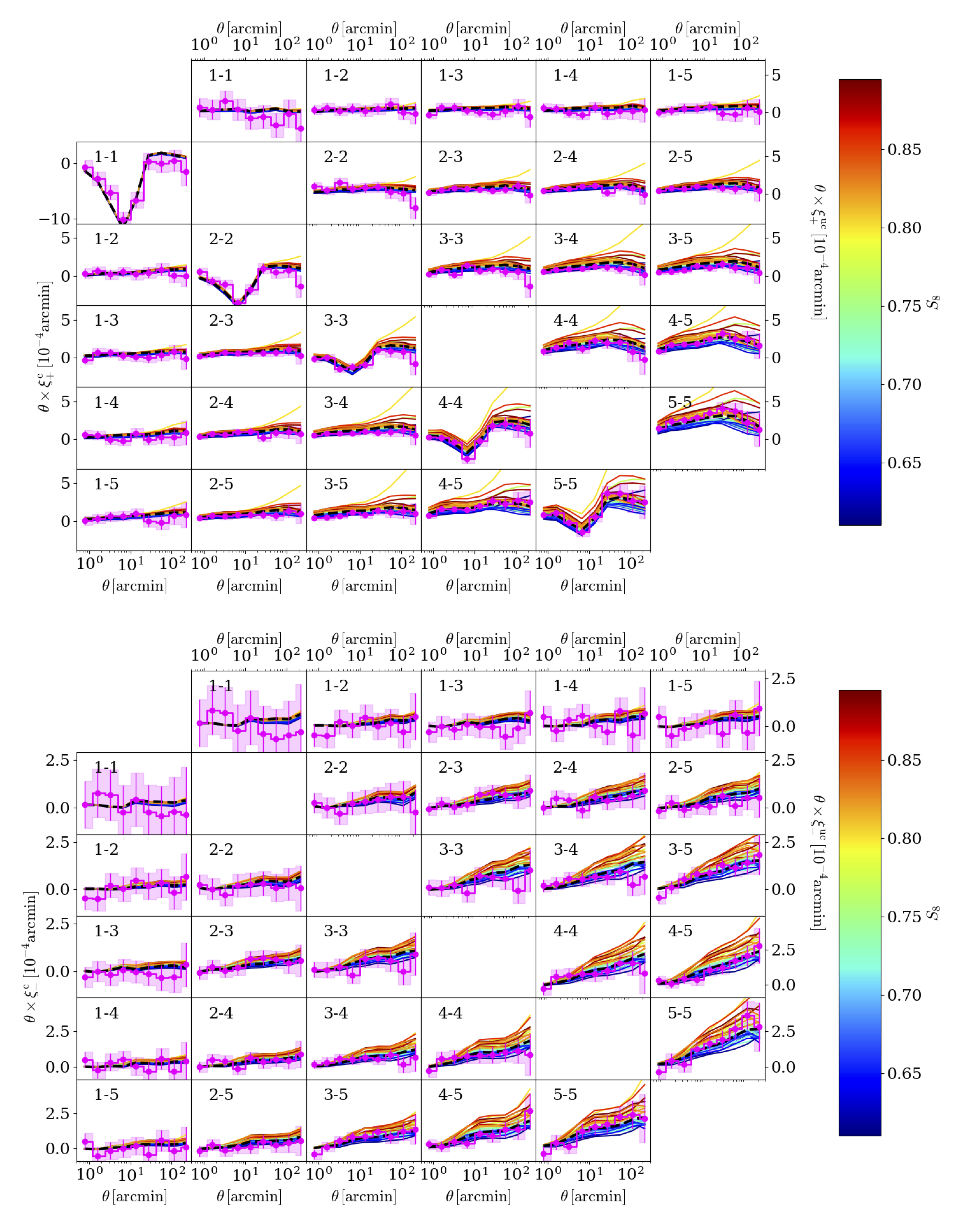}
\caption{The $\xi_+^{ij}(\theta)$ (upper block) and $\xi_-^{ij}(\theta)$ (lower block) scaled by $\theta \times 10^4$, with the clipped and unclipped measurements occupying the lower-left and upper-right corners of each block respectively. Each panel is annotated with the $i$-$j$ redshift bin combination. The {\sc cosmoSLICS} are shown colour-coded by their input $S_8$, the KiDS-1000 measurement by the magenta data data points, and the best-fit model by the dashed black line. The error bars on the KiDS-1000 measurement are the standard deviation estimated from the Covariance Set of simulations ({\sc SLICS}). Note that the $y$-axis differs for the clipped $\xi_+^{11}$ panel.}\label{fig:xi}
\end{center}
\end{figure*}

Using Eqn.~\ref{eqn:xi+-}, we measure unclipped, $\xi_\pm^{\rm uc}$, and clipped, $\xi_\pm^{\rm c}$ shear correlation functions\footnote{We henceforth neglect both the $\widehat{ }$ and `$ij$' redshift bin indicators when referencing $\xi_\pm$ (both clipped and unclipped) except in circumstances where inclusion is essential for clarity.} for the 18 tiles of the KiDS-1000 data before performing a weighted average across the tiles to obtain survey-wide measurements per redshift bin combination. In the case of our Cosmology Set of simulations, we obtain 50 realisations of the survey-wide $\xi_\pm^{\rm uc}$ and $\xi_\pm^{\rm c}$, which are then averaged per redshift bin and cosmology to produce measurements with suppressed sampling variance on which the emulators are trained. For our  IA Set of simulations, which are not infused with the KiDS-1000 footprint in the form of 18 tiles, the measurements are simply averaged across the 50 available realisations. 
 
From the $N=1,240$ realisations in our Covariance Set, we compute the combined covariance of the clipped and unclipped $\xi_\pm$ as
\begin{equation} \label{eqn:Cov}
\boldsymbol{C}(\theta_i, \theta_j) = \sum_k^{N}  \frac{ \left(\xi^k(\theta_i) - \overline{\xi}(\theta_i) \right) \left({\xi^k(\theta_j)} -  \overline{\xi}(\theta_j) \right)}{N-1} \,,
\end{equation}
where $\xi^k(\theta_i)$ denotes the concatenated $[\xi^{\rm c}_+, \xi^{\rm c}_-, \xi^{\rm uc}_+, \xi^{\rm uc}_-]$ measurement from the $k$'th mock survey realisation for $\theta_i \in [0.5,300] ^\prime$, and $\overline{\xi}(\theta_i)$ denotes its average across all realisations.

Figure \ref{fig:xi} shows the $\xi_+$ (upper block) and $\xi_-$ (lower block) shear correlation functions (scaled by $\theta \times 10^4$) measured from the Cosmology Set (colour-coded by $S_8$) and from KiDS-1000 (magenta data points) with the clipped in the lower-left corner and the unclipped in the upper-right of each block. The various panels, labelled $i$-$j$, present the 15 combinations of redshift bins. The error bars shown for the data measurement, depicting the expected statistical uncertainty for a single KiDS-1000 survey, are the square-root of the covariance-diagonal computed in Eqn.~\ref{eqn:Cov}. Also shown, by the dashed-black line, is the best-fit model obtained from our fiducial parameter inference settings discussed in Sec.~\ref{subsec:param_infer}. 

From the Cosmology Set measurements, the cosmological dependence of both the clipped and unclipped $\xi_\pm$, is most obvious in the higher redshift bins. This is expected, given the light from higher-redshift galaxies has been lensed by a larger volume of the Universe and hence displays a higher signal-to-noise ratio (SNR). The lower redshift bin measurements are dominated by galaxy shape noise and hence display limited dependence on $S_8$. For the higher-SNR bins, a clear positive trend between $\xi_\pm$ power and $S_8$ is observed for both the clipped and unclipped, also as expected, given larger $S_8$ values correspond to universes with greater matter clustering and hence larger lensing signal. This dependence, however, is diluted by the other changing cosmological parameters. A clear example of this is the yellow line, with $S_8 \approx 0.8$: an outlier in most redshift bins despite its intermediate $S_8$ value on account of it boasting the largest $\sigma_8$ (1.37) and lowest $\Omega_{\rm m}$ (0.1019) and hence, corresponding to an extreme universe with highly-clustered equal parts dark and baryonic matter. The {\sc cosmoSLICS} $\xi_\pm^{\rm uc}$ were compared with theoretical predictions in \cite{harnois-deraps/etal:2019} and found to be generally consistent at the level of a few percent on the scales at which the theory is valid.

Encouragingly, the KiDS-1000 data is bracketed by the measurements from the Cosmology Set across most of the redshift bins and angular separations for both the clipped and unclipped $\xi_\pm$. In some of the lower noise-dominated redshift bins, the Cosmology Set measurements, which feature reduced sample variance thanks to the averaging across multiple realisations, are instead bracketed by the relatively large uncertainties from the data. Whilst ``$\chi$-by-eye" assessments can be misleading for highly correlated data points, it is promising to note the apparent consistency between the mocks and KiDS-1000 data even in the absence of systematic biases being forward modelled into the {\sc cosmoSLICS} $\xi_\pm$.

The apparent consistency of the simulations and data is especially striking for the clipped $\xi_+$ (lower-left corner of the upper block in Fig.~\ref{fig:xi}), where the mocks recover the turning point at $\theta \simeq 6^\prime$ observed in the data. This decrement in power, examined in detail in G18, appears approximately at the scale of the smoothing kernel width and its depth is determined by the clipping threshold, $\kappa^{\rm c}$. The `clipping trough' appears most strongly in the auto-correlations of the $\xi_+^{\rm c}$, but is present to a lesser extent (a 10--20\% effect at $\theta \simeq 6^\prime$) in the cross-correlations as well. The depth of the clipping trough is also greater for the lower redshift bins. This is a consequence of the higher relative noise levels for low-redshift sources; as demonstrated in G18, the depth of the $\xi_+^{\rm c}(\theta \simeq 6^\prime)$ trough accentuates with increasing galaxy shape noise dispersion. This highlights the importance of accurately forward modelling the shape noise in the clipped measurement.

 \begin{figure*}
\begin{center}
\includegraphics[width=\textwidth]{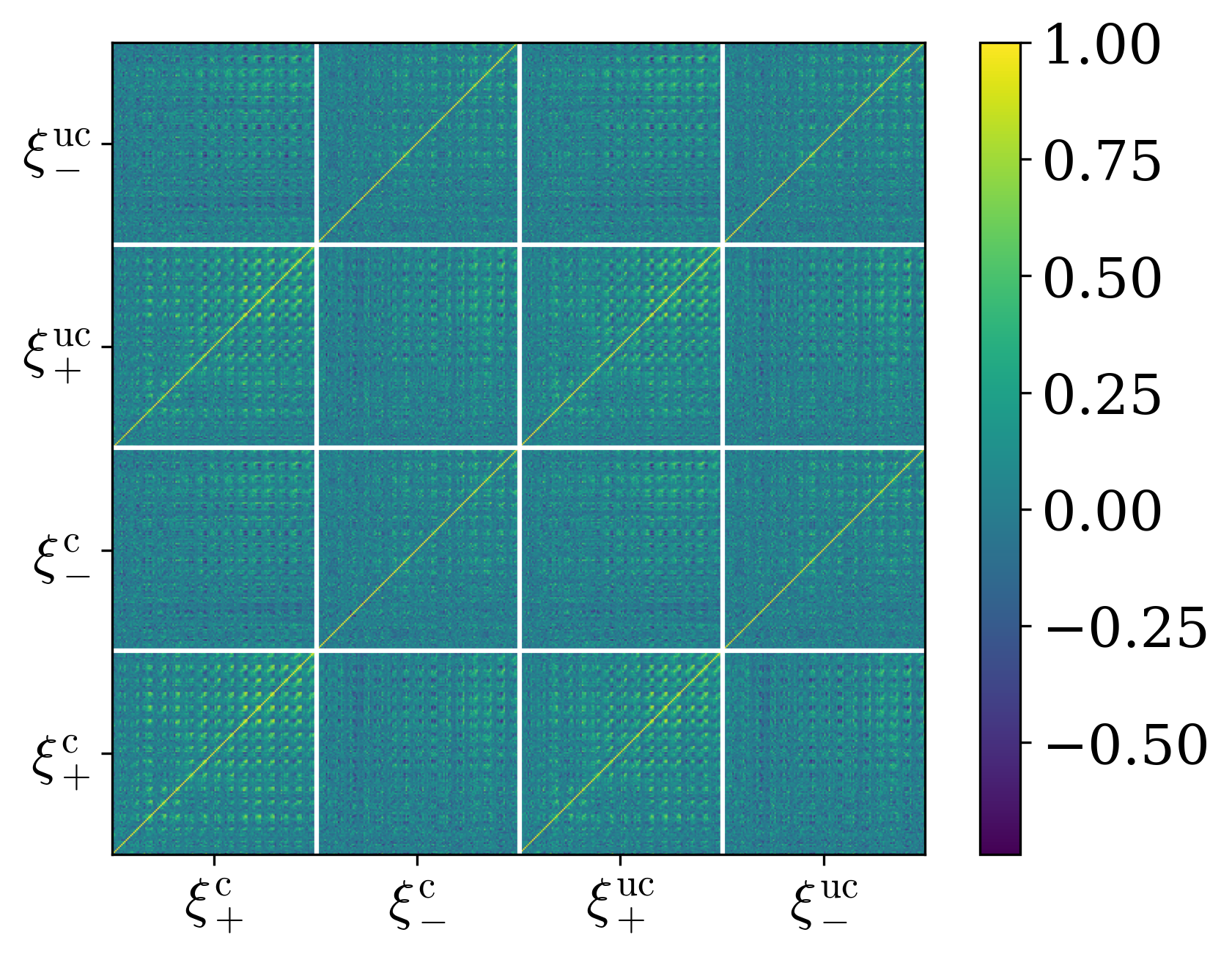}
\caption{The correlation coefficient matrix for the clipped, $\xi_\pm^{\rm c}$, and unclipped, $\xi_\pm^{\rm uc}$, statistics. Each block marked by the white lines consists of 135 elements (15 redshift bin combinations $\times$ 9 $\theta$ bins). }\label{fig:cov}
\end{center}
\end{figure*}  

The correlation coefficient matrix of the clipped and unclipped $\xi_\pm$, which is related to the covariance by
\begin{equation} \label{eqn:corr_coeff}
\boldsymbol{R}(\theta_i, \theta_j) = \frac{\boldsymbol{C}(\theta_i, \theta_j)}{\sqrt{\boldsymbol{C}(\theta_i, \theta_i) \boldsymbol{C}(\theta_j, \theta_j)}} \,,
\end{equation}
is shown in Figure \ref{fig:cov} with the clipped and unclipped $\xi_+$ and $\xi_-$ portions of the matrix delineated by white lines. Naturally, we see that there are strong correlations between the clipped and unclipped $\xi_+$, and correspondingly for the $\xi_-$. For example, the $(\xi_+^{\rm c}$, $\xi_+^{\rm uc})$ block has diagonal coefficients (i.e. measured at equivalent $\theta$ across the 15 redshift bin combinations) in the range 0.85--1.0 with a mean value of 0.96$\pm$0.03; similar metrics are derived for the $\xi_-$ portions. These coefficients are weakest at the small angular scales ($\xi_+ \left[ \theta \lesssim 10^\prime \right], \, \xi_- \left[ \theta \lesssim 50^\prime \right]$) since the clipping threshold and smoothing scale implemented were identified in G18 as appropriate for targeting the non-linear regime whilst leaving the linear scales mostly unaffected. It is these small deviations from pure correlation which allow for improvements in cosmological constraining power.

Large correlations within a covariance matrix have implications for the stability of matrix inversion. We verify the covariance matrix used in parameter inference is stable by measuring its condition number (the ratio of maximum and minimum eigenvalues), finding it to be on the larger side of reasonable ($\sim 2\times 10^8$). We therefore experiment with regularising the covariance by adding a small amount of jitter to the diagonal in order to lower the condition number to $1\times 10^7$; this has the impact of changing the diagonal elements by 0.6\% on average but is found to have negligible effect on the cosmological constraints.

\subsection{Emulator training} \label{subsec:emu_train}

Effective parameter inference requires accurate model predictions as a function of \textit{arbitrary} values of the cosmological and nuisance parameters, facilitating computation of the likelihood at each step in the parameter-space-sampling chain. In order to interpolate between the simulation nodes, at which the clipped and unclipped $\xi_\pm$ are known precisely, we use Gaussian process regression (GPR) emulators (implemented with the {\sc scikit-learn} package). GPR emulators use a training set to fit the free hyperparameters of a kernel describing the covariance between the predictions in the model parameter space. We adopt a common choice for the kernel - the squared exponential (or Gaussian) - which features $d$ hyperparameters describing the per-dimension correlation length of the learnt statistic, plus an overall kernel amplitude. We train emulators to learn the cosmological dependence of the clipped and unclipped $\xi_\pm$ in each of the 15 redshift bin combinations separately. Hence, each of these $15 \times 2$ emulators features $d+1$ hyperparameters, where $d=4$ for the $[\Omega_{\rm m}, S_8, h, w_0]$ parameter space.

 \begin{figure*}
\begin{center}
\includegraphics[width=0.8\textwidth]{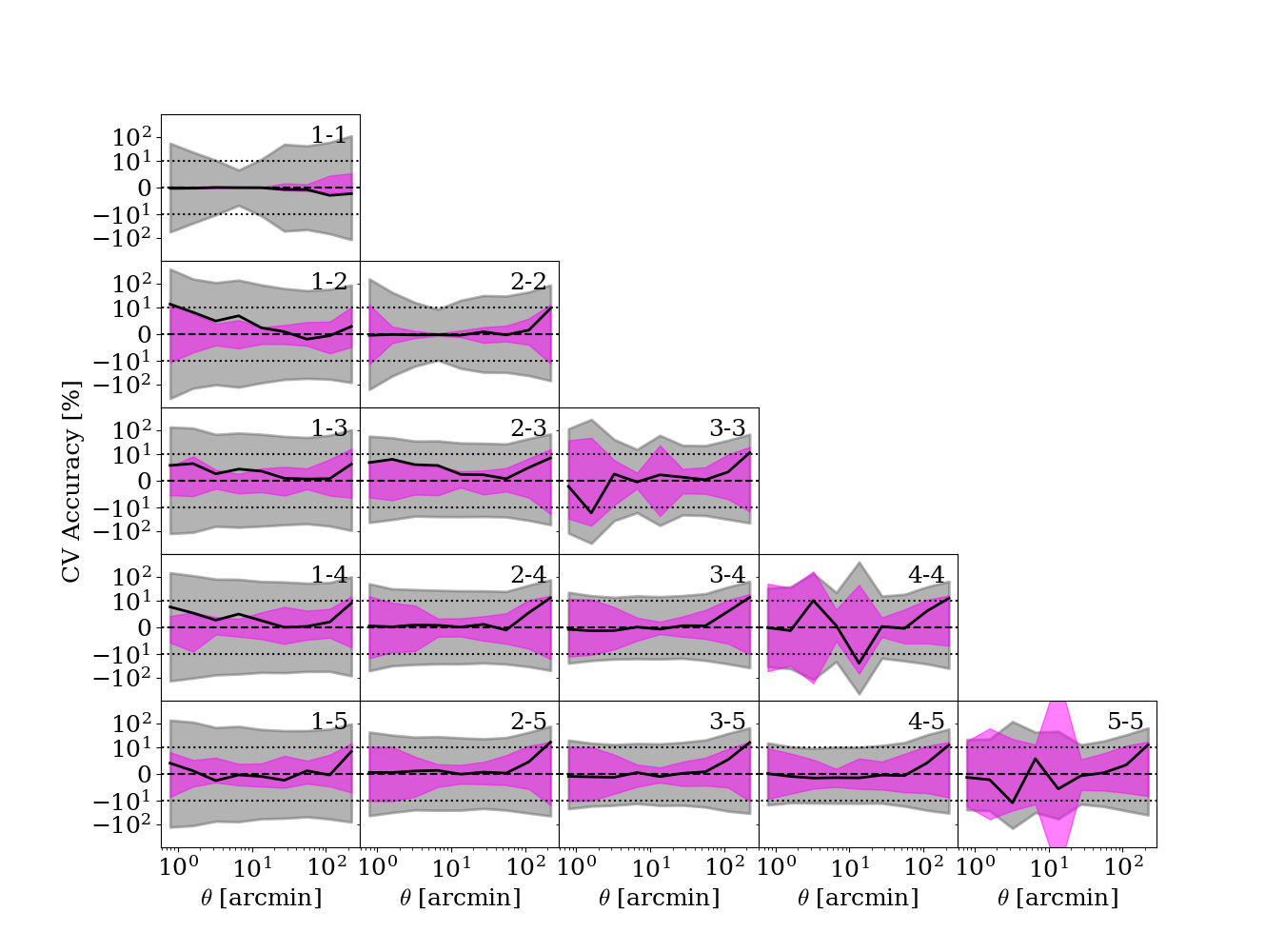}
\caption{The accuracy of the cosmological emulators of $\xi_+^{\rm c}(\theta)$, assessed via cross-validation (CV) represented by the magenta band which is twice the range spanned by 68\% of the measurements across the training nodes. The grey band shows twice the statistical uncertainty for the KiDS-1000 survey (grey band) and the solid black lines show the CV-accuracy for the fiducial cosmology.
The panels represent the different redshift bin combinations and the horizontal dashed lines indicate where the scale changes from linear to log ($\pm 10\%$).  }\label{fig:cvacc_clip}
\end{center}
\end{figure*} 

Our emulation strategy broadly follows that of the original {\sc cosmoSLICS} emulator, demonstrated in \cite{harnois-deraps/etal:2019} to achieve 1--5\% accuracy for the unclipped $\xi_\pm$ in a single broad redshift bin across a wide range of cosmological parameter space and angular scales. This consists of transforming the clipped and unclipped $\xi_\pm$ to narrow the dynamic range to $\mathcal{O}(1)$ before performing a principal component analysis (PCA) to compress the transformed statistics into $n_\phi$ orthogonal basis functions. Previously, a logarithmic transform was used, but as shown in Fig.~\ref{fig:xi}, this is an inappropriate choice for the \textit{tomographic} unclipped and clipped $\xi_\pm$ which feature many (in some cases noise-driven) negative values across the data vector. We therefore scale the $\xi_\pm$ by $\theta \times 10^4 \, [{\rm arcmin}]$ (the scaling shown in Fig.~\ref{fig:xi}) and for the PCA, we chose $n_\phi$ equal to the number of $\theta$ bins per statistic. The emulators in each redshift bin are then trained to predict the weights of the nine principal components. We include the PCA for most of the parameter inference chains ran in our analysis, but we also find that omitting the PCA has no impact on the final cosmological constraints.

In order to verify the accuracy of our cosmological emulators, we perform a `leave-one-out cross validation' (CV) analysis. This consists of iteratively omitting each node from the training and comparing the emulators' predictions for the missing node to the true (simulated) measurement. CV gives a conservative estimate of the emulation accuracy across the parameter space given all predictions are derived from a slightly reduced training set size. In Figure \ref{fig:cvacc_clip}, the percentage differences between the predicted and true $\xi_+^{\rm c}(\theta)$ obtained through CV are represented with the magenta band. Specifically this band is double the range spanned by 68\% of the accuracy measurements from the nodes of the training set but omits the most extreme nodes in each of the 4 cosmological dimensions (nodes where the emulator is required to extrapolate) as these give a biased measure of the accuracy in the main body of the parameter space.

For the majority of redshift bins and angular scales, the uncertainty in the emulators' predictions are subdominant to the statistical uncertainty for KiDS-1000 (the grey band shows twice the uncertainty); this is calculated as the standard deviation from our Covariance Set normalised by the fiducial cosmology\footnote{One could also have normalised by the KiDS-1000 data measurement itself, which manifests in a noisier version of the grey band pictured.} and expressed as a percentage. The fiducial cosmology measurement is shown by the solid black line. As this node lies close to the centre of the parameter space, it can be regarded as representative of the emulator accuracy for values of the cosmological parameters of greatest interest (generally a few \%). Similar and slightly improved CV-accuracies are seen for the $\xi_-^{\rm c}$ and $\xi\pm^{\rm uc}$ respectively.

These results suggest that in general the emulators are sufficiently well conditioned given the statistical power in the KiDS-1000 survey, but to avoid any emulator inaccuracies biasing the cosmological constraints, we fold the residual emulator uncertainty assessed via CV into the covariance. Specifically, we follow \cite{harnois-deraps/etal:2024} and inflate the diagonal of the covariance as follows:
\begin{equation} \label{eqn:Cov_emu}
\mathbf{C}^{\rm emu} = \mathbf{C} + {\rm diag} \left\langle {\left( \xi^{\rm emu}-\xi^{\rm sim} \right)}^2 \right\rangle .
\end{equation}

 \begin{figure*}
\begin{center}
\includegraphics[width=\textwidth]{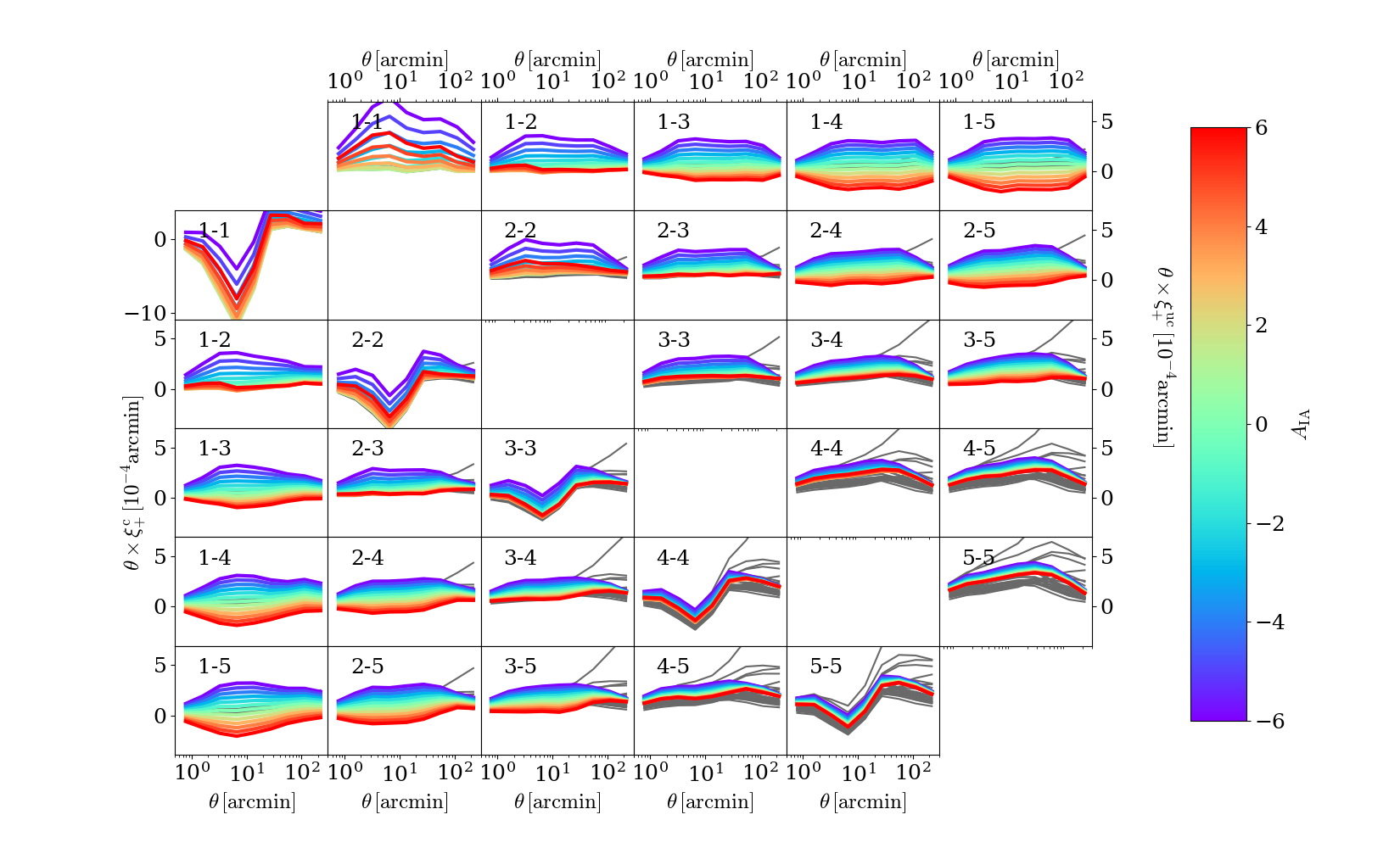}
\caption{The fiducial cosmology $\xi_+$ (scaled by $\theta \times 10^4 \, [\rm arcmin]$) contaminated with intrinsic alignments of various strengths (colour-coded by $A_{\rm IA}$), with the clipped and unclipped measurements occupying the lower-left and upper-right corners respectively, relative to the (systematics-free) Cosmology Set (grey lines; previously shown in Fig.~\ref{fig:xi}).}\label{fig:clip_xipm_sysIA}
\end{center}
\end{figure*} 

\noindent Here $\mathbf{C}^{\rm emu}$ is the covariance matrix including the emulation error, $\mathbf{C}$ is the matrix measured from our Covariance set of simulations, whilst $\xi^{\rm emu}$ and $\xi^{\rm sim}$, respectively, stand in for the clipped and unclipped $\xi_\pm$ obtained from the emulators in CV and from the simulated ground truth. The averaging of the squared differences between the predictions and the truth is done across $n$ nodes of the training set considered to be representative of the parameter space volume of greatest interest. We experimented with different values for $n$, ranging from using all nodes except for those at the outermost edge of the parameter space \citep[the approach adopted by][which in practice corresponds to $n=20$]{harnois-deraps/etal:2024}, selecting the $n=12$ or 6 nodes closest to the fiducial cosmology, to neglecting the emulator error altogether ($n=0$). We find that all definitions of the emulator error return consistent cosmological and systematic constraints, but those derived from the covariance neglecting the emulation error return a poor goodness-of-fit (evaluated via a posterior predictive $p$-value). All values of $n>0$ which we tried return very similar constraints and goodness-of-fit measures however, and so we adopt the $n=12$ definition of the emulation error for our fiducial parameter inference settings. The results of these tests, including an alternative definition based on the cosmology-dependent Gaussian process covariance, are presented in Appendix \ref{sec:appendix_emu_err}. We also explored including in the covariance the errors from our emulators of systematic biases, discussed in Sec.~\ref{subsec:mod_sys},  but found the constraints were unchanged due to the sub-dominance of the systematic emulator errors to those of the cosmological emulators. We therefore neglect the former emulator errors.

\subsection{Modelling systematics} \label{subsec:mod_sys}

For modelling the systematic impacts on the clipped and unclipped $\xi_\pm$ using the IA, BF and photo-$z$ training sets (see Sec.~\ref{subsec:sims}), we use GP emulators for cases where the number of training nodes is large and/or the interpolation multi-dimensional, and simple linear models otherwise. 

For example, since the strength of IAs in our simulations can be modified in post-processing via a simple rescaling of the $A_{\rm IA}$ parameter, we are able to generate an arbitrary number of training predictions for this systematic, subject only to the computational expense of performing the mass mapping for all realisations and redshift bins.  We therefore generate $\xi_\pm^{\rm c/uc}$ for 13 IA nodes uniformly sampling the prior space adopted in the A21 cosmic shear analysis, $A_{\rm IA} \in [-6,6]$. Following the fiducial emulation strategy applied to the Cosmology training set, GP emulators are then trained to predict the IA biases, 
\begin{equation}
B_{\rm IA} = \xi_{A_{\rm IA}} - \xi_{(A_{\rm IA}=0)} \,,
\end{equation}
where $\xi$ denotes the clipped and unclipped $\xi_\pm$ in each redshift bin.

Figure \ref{fig:clip_xipm_sysIA} shows the impact of IAs as a function of $A_{\rm IA}$ (various colours) on the fiducial cosmology $\xi_+$ relative to the cosmology training set (grey lines). We see the IA contribution to $\xi_+^{\rm c}$ exceeds the impact of changing cosmological parameters in the noise-dominated low-redshift bins, where the cosmological signal is weak. The IA impact is also relatively high for the cross-correlations; this is caused by the `gravitational-intrinsic' (GI) correlations between galaxies at different redshifts arising from the coupling of a foreground galaxy with its local density, which in turn contributes to the lensing distortion of a background galaxy \citep{hirata/seljak:2004}. The GI term is typically larger than the `intrinsic-intrinsic' (II) correlations between the shapes of galaxies at the same redshift due to their formation within a shared environment; hence why the auto-correlations are impacted less by IAs relative to changing cosmology. 

The IA Set also allow us to investigate the potential biases from \textit{source-lens clustering} (SLC): the preferential sampling of the shear field at higher values due to galaxies clustering along overdense sight-lines. \cite{gatti/etal:2024} showed that the impact of SLC varies across HOWLS but can be significant (e.g. in the case of wavelet phase harmonics, where a $6.5\sigma$ detection of SLC was detected in the signal from DES Year 3 data). We investigated the bias from SLC to the clipped (and unclipped) $\xi_\pm$, finding it to be negligible and the constraints unchanged when the amplitude of the SLC is included as an additional nuisance parameter. The details of and results from these tests are presented in Appendix \ref{sec:appendix_slc}. We are therefore neglect the effects of SLC in the primary analysis. 

For the Photo-$z$ Set of simulations, we use a combination of GP emulators and, following \cite{harnois-deraps/etal:2021,harnois-deraps/etal:2024}, linear interpolation to model the impact of the  biases to $\xi_\pm^{\rm c/uc}$ from shifting the mean of the $n(z)$ in each of the five tomographic bins by a per-bin nuisance parameter, $\Delta z_i$ (with $i \in [1,5]$). Specifically, since the biases to the cross correlations are a function of \textit{two} nuisance parameters, $\boldsymbol{\Delta z_{ij}} = [\Delta z_i, \Delta z_j]$ ( where $i \neq j$), we use GP emulators to perform the 2D interpolation between $4 \times 4$ simulation nodes given by all combinations of the four shifts, $\boldsymbol{\Delta z} = \left( -0.898, -0.229, 0.262, 0.8254 \right) \boldsymbol{\sigma_z}$. The uncertainty in the bias, $\boldsymbol{\sigma_z}$, varies slightly between the five redshift bins (see Sec.~\ref{subsec:sims}), meaning the nodes are in slightly different locations in the $[\Delta z_i, \Delta z_j]$ parameter space for different redshift bin combinations (e.g. $[\Delta z_1, \Delta z_2]$ vs $[\Delta z_1, \Delta z_3]$). Given that we always use bin-specific emulators, however, this node shifting poses no confusion to the training. 

As with the IA Set, the emulators predict the bias from shifting $n(z)$:
\begin{equation} \label{eqn:bias_dz}
B_{\boldsymbol{\Delta z_{ij}}} = \xi_{\boldsymbol{\Delta z_{ij}}} - \xi_{(\boldsymbol{\Delta z_{ij}=0})}.
\end{equation} 
Unlike the IA Set, however, we have no baseline Photo-$z$ simulation with $\Delta z=0$ and with the exact same noise properties\footnote{We investigated using the fiducial node in the (systematics-free) Cosmology Set of simulations for the $\Delta z=0$ baseline measurement, but found that the difference in the noise levels of the Photo-$z$ and Cosmology Sets (due to the different number of realisations) led to non-trivial trends in the biases as a function of $\boldsymbol{\Delta z_{ij}}$, which would have been ineffectively modelled with either the GP emulators or linear models.}. We therefore perform an initial training on the Photo-$z$ Set to predict the clipped and unclipped $\xi_{(\boldsymbol{\Delta z_{ij}=0})}$ (following the fiducial emulation strategy outlined above) which is then used to compute the biases, $B_{\boldsymbol{\Delta z_{ij}}}$ via Eqn.~\ref{eqn:bias_dz}, before retraining the emulators to predict this quantity.

Given that we have only 16 nodes to model the dependence of $B_{\boldsymbol{\Delta z_{ij}}}$ on two nuisance parameters, to further improve the interpolation accuracy we use separate emulators for each of the nine $\theta$ bins per redshift bin and statistic ($\xi_\pm^{\rm c/uc}$). This represents a divergence from the strategy with our cosmology and IA emulators, where one emulator made predictions across all angular scales for a given statistic and redshift, and increases the computational expense of sampling the nuisance parameter space. However, this means that the $d+1$ hyperparameters of the GP kernel (where $d=2$ for $\boldsymbol{\Delta z_{ij}}$) need only model the single $B_{\boldsymbol{\Delta z_{ij}}}(\theta)$ value, rather than a nine-bin function, leading to greater overall accuracy.

The photo-$z$ biases to the \textit{auto}-correlations are, conversely, modelled as a function of only one nuisance parameter with our four nodes in  the $\Delta z_i$ parameter space. Owing to the one-dimensional nature of the interpolation and the small number of nodes, we follow \cite{harnois-deraps/etal:2021,harnois-deraps/etal:2024} and fit a linear model to $\theta \times \xi_{\Delta z_i}$ per redshift and $\theta$ bin for the clipped and unclipped $\xi_\pm$ separately in order to predict $\xi_{(\Delta z_i=0)}$. We then compute the bias to the auto-correlation as 
\begin{equation} \label{eqn:bias_dz2}
B_{\Delta z_i} = \xi_{\Delta z_i} - \xi_{(\Delta z_i=0)} \,. 
\end{equation} 
Finally, we re-fit the linear relation to the bias parameters, which changes only the $y$-axis intercept and not the gradient of the initial model. 

\begin{figure}
\begin{center}
\includegraphics[width=0.5\textwidth]{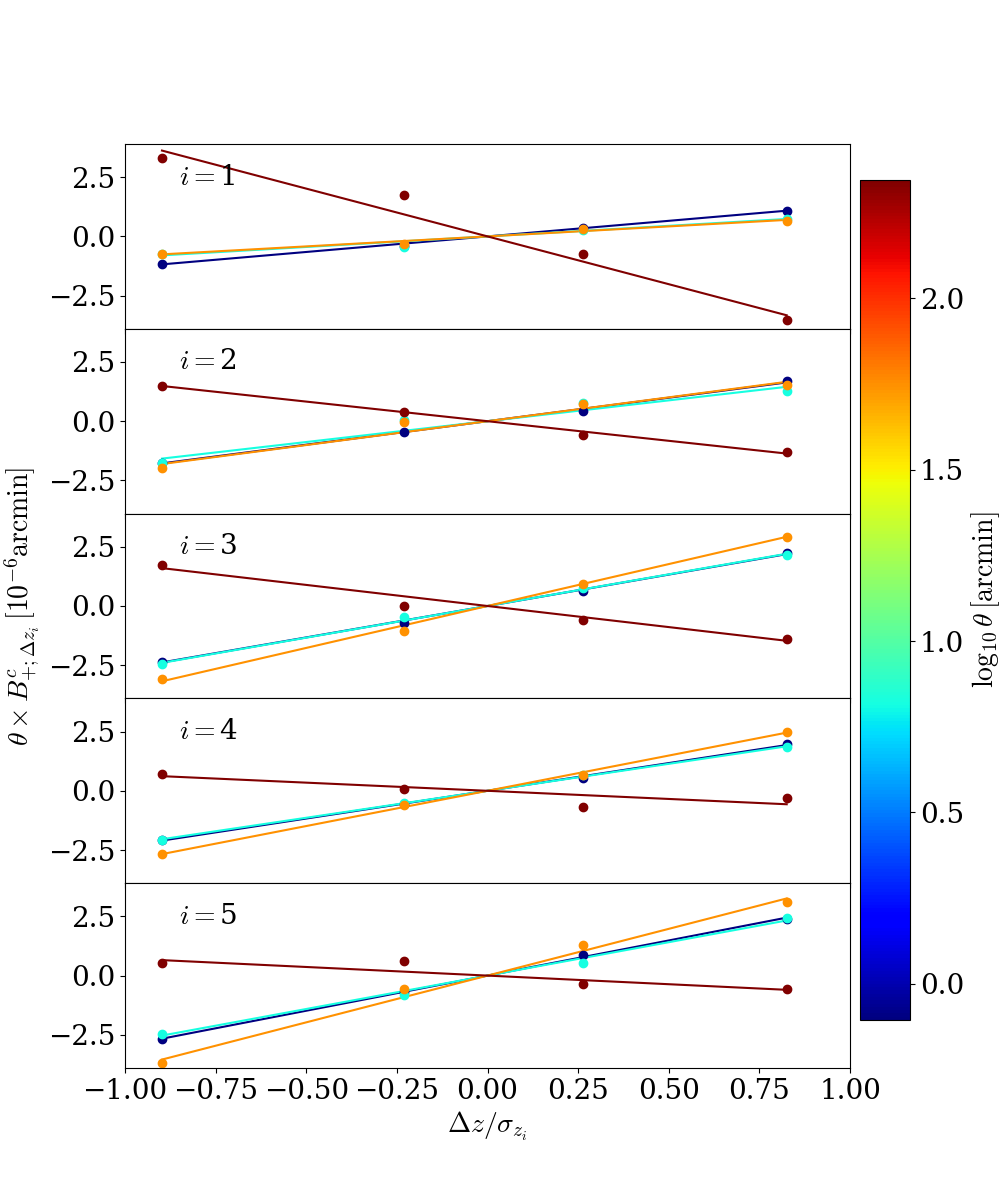}
\caption{Biases, $B^c_{+; \Delta z_i}$ (Eqn.~\ref{eqn:bias_dz2}) in four $\theta$ bins (as shown by the colour bar) to the $\xi_+^{\rm c}$ auto-correlations in the five tomographic bins ($i \in [1,5]$; upper to lower panels respectively) as a function of the shift to the mean of the redshift distribution, $\Delta z / \sigma_i$, where $\sigma_i$ is the uncertainty on the bias in each bin \citep{giblin/etal:2021}. The data points are measured from our Photo-$z$ Set of mocks (see Sec.~\ref{subsec:sims}) and the lines are the linear models.  Note that the biases are scaled by $\theta \times 10^6 \, [\rm arcmin]$ and hence are  $\sim $100$\times$ smaller than the IA-biases depicted in Fig.~\ref{fig:clip_xipm_sysIA}.}\label{fig:clip_xip_sys_dz}
\end{center}
\end{figure}

Figure \ref{fig:clip_xip_sys_dz} shows the biases and associated linear models for four $\theta$ bins of the $\xi_+^{\rm c}$ auto-correlations (as indicated by the colours) with the five panels corresponding to the  five tomographic bins ($i \in [1,5]$). We see that the biases from shifts to the redshift distributions are well-approximated with a linear relation across a wide range of redshifts and angular separations, and that there are no obvious correlations between $B_{\Delta z_i}$ and $\theta$, justifying the bin-wise modelling. We do not propagate uncertainties on the models for $B_{\Delta z_i}$ to parameter inference as the photo-$z$ biases are $\sim$100$\times$ smaller than the biases from intrinsic alignments depicted in Fig.~\ref{fig:clip_xipm_sysIA} (note there is a factor of $10^2$ difference in the scaling between these two figures) and smaller still than the errors in the cosmological emulators. Hence, photo-$z$ biases represent a subdominant source of systematics in this analysis.

 \begin{figure*}
\begin{center}
\includegraphics[width=\textwidth]{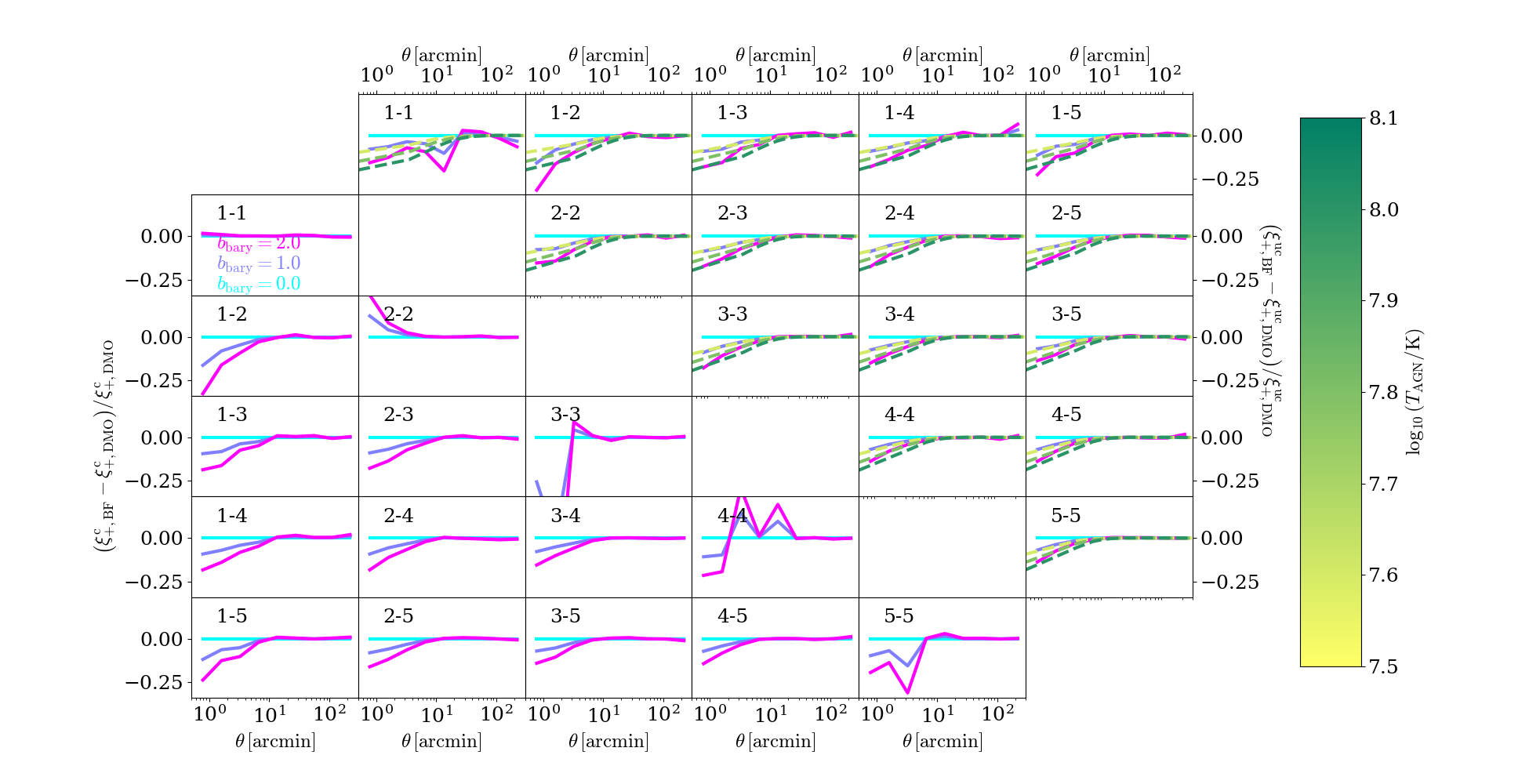}
\caption{The fractional impact of baryonic feedback (BF) on the clipped (lower left) and unclipped (upper right) $\xi_+$ measured relative to the dark-matter-only (DMO) {\sc magneticum} simulation ($b_{\rm bary}=0.0$; cyan line). The BF-contaminated {\sc magneticum} simulation ($b_{\rm bary}=1.0$) is shown by the purple line along with the extrapolated impact ($b_{\rm bary}=2.0$; magenta) permitted by the prior.  For comparison, on the $\xi_+^{\rm uc}$ panels we also show the fractional difference between {\sc hmcode} predictions with $\log_{10} \left( T_{\rm AGN}/{\rm K} \right)=[7.6, 7.8, 8.0]$ (light to dark green dashed lines respectively) and the {\sc hmcode} prediction with no BF \citep{mead/etal:2021}.}\label{fig:xip_bary}
\end{center}
\end{figure*}

Finally, we also model the biases from baryons following \cite{harnois-deraps/etal:2021,harnois-deraps/etal:2024} with a per-redshift-bin and per-$\theta$ linear model as a function one nuisance parameter, $b_{\rm bary}$, which takes the values of 1.0 and 0.0 in our baryon-contaminated and dark-matter-only mocks respectively. Figure \ref{fig:xip_bary} shows the fractional impact of BF on the clipped (lower corner) and unclipped (upper corner) $\xi_+$ for $b_{\rm bary}=[0.0, \, 1.0, \, 2.0]$ (cyan, purple and magenta respectively) measured from {\sc magneticum}, where the $b_{\rm bary}=2.0$ result is derived by simply doubling the bias seen in the $b_{\rm bary}=1.0$ simulation. For comparison, we also show {\sc hmcode}  $\xi_+^{\rm uc}$ predictions \citep[calibrated on BAHAMAS;][]{mead/etal:2021} for varying $\log_{10}\left( T_{\rm AGN} / {\rm K} \right)$, as indicated by the colour bar, via the dashed lines in the upper-corner panels\footnote{{\sc hmcode} of course does not provide $\xi_+^{\rm c}$ predictions, and hence these lines do not feature on the lower corner panels.}. The impact of BF on the $\xi_+^{\rm uc}$ ($\sim$5--8\% suppression on scales of a few arcmin) from {\sc magneticum} (purple) is closest to the $T_{\rm AGN}=10^{7.6}{\rm K}$ (light green) prediction, whereas our extrapolated $b_{\rm bary}=2.0$ (magenta) generally falls between the $T_{\rm AGN}=10^{7.8} \, {\rm K}$ and $10^{8.0} \, {\rm K}$ results. By allowing for $b_{\rm bary}$ to vary in this range in our modelling, therefore, generally approximates the prior range assumed for $T_{\rm AGN}$ in other cosmic shear analyses \citep[see, for example,][]{DES/KiDS:2023}.

The effect of BF on the $\xi_+^{\rm c}$ is similar to the unclipped for the cross-correlations, where the characteristic clipping trough at $\theta \simeq 6^\prime$ is less prominent. For the auto-correlations, however, the behaviour of BF as a function of $b_{\rm bary}$ is complicated by clipping: stronger BF generally acts to reduce the matter power on small scales which can lead to \textit{less} structure being clipped and consequently \textit{more} power on small scales, contrary to what is seen in the unclipped statistic. The increases in the BF-contaminated $\xi_+^{\rm c}(5^\prime \lesssim \theta \lesssim 10^\prime)$ relative to the dark-matter-only measurement, seen in the auto-correlations from redshift bins 2--5, are likely caused by this effect: fewer high-density structures exist on these scales in the baryon mocks, and less of the field is clipped accordingly. This effect is redshift dependent, however, and so we see the clipping trough in the auto-correlations changing in a non-trivial way as a function of $b_{\rm bary}$ and redshift bin combination, whereas, for the cross-correlations, the impact mostly mirrors those of the unclipped statistic.

\subsection{Parameter inference} \label{subsec:param_infer}

The posterior probability of certain values of the cosmological and nuisance parameters, $\boldsymbol{\pi}=\left[ \Omega_{\rm m}, S_8, h, w_0, A_{\rm IA}, \boldsymbol{\Delta z}, b_{\rm bary} \right]$, being the truth given some observed data, $\boldsymbol{d}$, is given by Bayes' theorem,
\begin{equation}
p(\boldsymbol{\pi} | \boldsymbol{d} ) = \frac{\mathcal{L}(\boldsymbol{d} | \boldsymbol{\pi} ) p(\boldsymbol{\pi})}{E(\boldsymbol{d})} \,,
\end{equation}
where $p(\boldsymbol{\pi})$ is the prior probability on the parameters and $E(\boldsymbol{d})$ is the Bayesian evidence which normalises the integral of the posterior over all values of $\boldsymbol{\pi}$ to unity. A Gaussian form is often assumed for the likelihood, 
\begin{equation} \label{eqn:Lhd_Gauss}
\mathcal{L_G}(\boldsymbol{d}|\boldsymbol{\pi}) =  \frac{1}{\sqrt{(2\pi)^D |\boldsymbol{\Sigma}|}} \exp \left( -\frac{1}{2} \left[\boldsymbol{d} - \boldsymbol{m}(\boldsymbol{\pi}) \right]^\intercal \boldsymbol{\Sigma}^{-1} \left[\boldsymbol{d} - \boldsymbol{m}(\boldsymbol{\pi}) \right] \right) \,,
\end{equation}
where $\boldsymbol{m}(\boldsymbol{\pi})$ is the model prediction (given by the emulators and linear fits in this case), $D$ is its length and $\boldsymbol{\Sigma}$ is the \textit{true} covariance matrix. Whilst the covariance estimated from our simulations, $\boldsymbol{C}$ (Eqn.~\ref{eqn:Cov}), is an unbiased estimate of $\boldsymbol{\Sigma}$, shot noise in the realisations means the same cannot be said of the inverse: $\boldsymbol{C}^{-1}$ is not an unbiased estimate of $\boldsymbol{\Sigma}^{-1}$. To mitigate potential biases in the best-fit $\boldsymbol{\pi}$ arising from this fact, \cite{hartlap/etal:2007} propose scaling the simulated inverse-covariance by a correction factor,
\begin{equation}
\widehat{\boldsymbol{C}^{-1}} = \frac{N-D-2}{N-2}\boldsymbol{C}^{-1} \,,
\end{equation}
where $N=1,240$ is the number of simulation realisations and the data vector length, $D$, is 540 and 450 before and after angular scale cuts respectively for the combined clipped and unclipped $\xi_\pm$. A more robust approach, however, given by \cite{sellentin/heavens:2016}, is to marginalise over the true covariance matrix; in practice, this means adopting a multivariate $t$-distribution for the likelihood in place of a Gaussian:
\begin{equation}
\mathcal{L}_t = \frac{\Gamma\left( \frac{N}{2} \right)}{ [\pi(N-1)]^{D/2} \, \Gamma\left( \frac{N-D}{2} \right) } 
\frac{|\boldsymbol{C}|^{-1/2}}{ \left( 1 + \frac{\left[\boldsymbol{d} - \boldsymbol{m}(\boldsymbol{\pi}) \right]^\intercal \boldsymbol{C}^{-1} \left[\boldsymbol{d} - \boldsymbol{m}(\boldsymbol{\pi}) \right]}{N-1} \right)}  \,,
\end{equation}
where $\Gamma(x)$ is the Gamma function. We ran chains using both forms for the likelihood under our fiducial parameter inference settings (choice of scale cuts, treatment of systematics and emulator error, explained further in this section) and found nearly identical results. We therefore adopt the Gaussian likelihood with the \cite{hartlap/etal:2007} covariance scaling for all constraints presented in this work. Our covariance is scaled after incorporating the emulator errors following Eqn.~\ref{eqn:Cov_emu}.

An implicit assumption whilst using the Gaussian and $t$-distributed likelihoods is that the data is itself Gaussian distributed. We confirm this is valid by assessing, via Kolmogorov-Smirnov (KoS) tests, the consistency between the Covariance Set clipped and unclipped $\xi_\pm^{ij}(\theta)$ realisations, in each redshift and $\theta$ bin, and Gaussian distributions with the same mean and variance. Adopting a 5\% threshold for the $p$-values, we find that the vast majority (97\%) of the 540 elements in the $\xi_\pm^{\rm c/uc}$ data vector pass the KoS test. This low level of outliers, therefore,  justifies our use of the Gaussian and $t$-distributed likelihoods for parameter inference. 

For exploring the 11-dimensional (four cosmological and seven nuisance) parameter space, we use the {\sc nautilus} \citep{lange:2023} sampler which employs importance nested sampling aided by efficient proposal functions derived by deep learning. We also found very similar results with {\sc emcee} \citep{foreman-mackey/etal:2013} which exercises a traditional Metropolis-Hastings Markov Chain Monte Carlo (MCMC).

Our fiducial parameter inference settings use the same scale cuts as A21: $\theta \in [0.5^\prime,300^\prime]$ for $\xi_+$ and $\theta > 4^\prime$ for $\xi_-$ (both clipped and unclipped). We also adopt the same priors as the {\sc cosmoSLICS}-based analysis by \cite{harnois-deraps/etal:2024} with the exception that a slightly wider $A_{\rm IA}$ range was used, matching that of A21. The priors are shown in Table \ref{tab:priors} and are broadly set by the range of input parameters to the simulations used in training the emulators. The mean shifts to the redshift distributions in the five tomographic bins, $\Delta z_i$, are correlated. Consequently, following A21 and \cite{harnois-deraps/etal:2024}, these are sampled from a multivariate Gaussian distribution with mean $\boldsymbol{\mu}_z = [0.0, 2.0, 1.3, 1.1, 0.6]\times 10^{-2}$ and covariance,

\begin{equation} \label{eqn:cz}
\boldsymbol{C}_z = 
\begin{bmatrix}
11.20 & 2.600   & 1.562  & 0.056  & 0.622   \\
2.600 & 12.78   & 4.081  & -1.692 & -0.2140  \\
1.562 & 4.081   & 13.81  & -1.139 & 0.525    \\
0.056 & -1.692  & -1.139 & 7.551  & 3.054     \\
0.622 & -0.2140 & 0.525  & 3.054  & 9.496      \\
\end{bmatrix} \times 10^{-5} \,,
\end{equation} 
where the order of elements is such that the upper-left and lower-right corners correspond to the first and fifth redshift bins respectively. The $\boldsymbol{\mu}_z$ and $\boldsymbol{C}_z$ values were obtained by measuring in MICE2 simulations the average offsets and covariance between the true $n(z)$ and those obtained with the KiDS-1000 redshift calibration \citep{hildebrandt/etal:2021}. The way this is implemented in sampling the parameter space is as follows: five uncorrelated $\Delta z_i$ are drawn from the priors which are then rotated via a matrix multiplication with the Cholesky decomposition of $\boldsymbol{C}_z$ to generate five correlated $\Delta z_i$ parameters. These are then are fed to the emulator/linear models.

\begin{table}
  \centering
    \caption{Priors used in inference of the cosmological and nuisance parameters (upper and lower sections respectively). All priors are flat except for the five offsets to the means of the tomographic redshift distributions, $\boldsymbol{\Delta z}$. As correlated nuisance parameters, these are drawn from a multivariate Gaussian with mean $\boldsymbol{\mu}_z$ and covariance $\boldsymbol{C}_z$ given in Eqn.~\ref{eqn:cz}. The boundaries of the flat priors are set to the ranges of our training simulations with the exception of $b_{\rm bary}$, where extrapolation to twice the strength of BF in {\sc magneticum} is permitted.} \label{tab:priors}
  \begin{tabular}{lr}
Parameter    & Prior \\
  \hline
$\Omega_{\rm m}$ & $[0.10, 0.55]$ \\
$S_8$ & $[0.6, 0.9]$ \\
$h$ & $[0.60, 0.82]$ \\
$w_0$ & $[-2.0, -0.5]$ \\ \hline
$A_{\rm IA}$ &  $[-6, 6]$ \\
$\boldsymbol{\Delta z}$ &  $\mathcal{G}(\boldsymbol{\mu}_z, \boldsymbol{C}_z)$ \\
$b_{\rm bary}$ & $[0, 2]$ \\
  \hline
  \end{tabular}
\end{table}

%% file: Section_4_Results.tex
\section{Results}
\label{sec:results}

Prior to deriving constraints from the KiDS-1000 data, we validated our emulator-based modelling and fiducial parameter inference settings with both systematics-contaminated mock data, and external mock data to the cosmology training set. We find that unbiased cosmological and systematic constraints are recovered in all analyses; the results of these tests are presented in Appendix \ref{sec:appendix_fakedata}.

 \begin{figure*}
\begin{center}
\includegraphics[width=\textwidth]{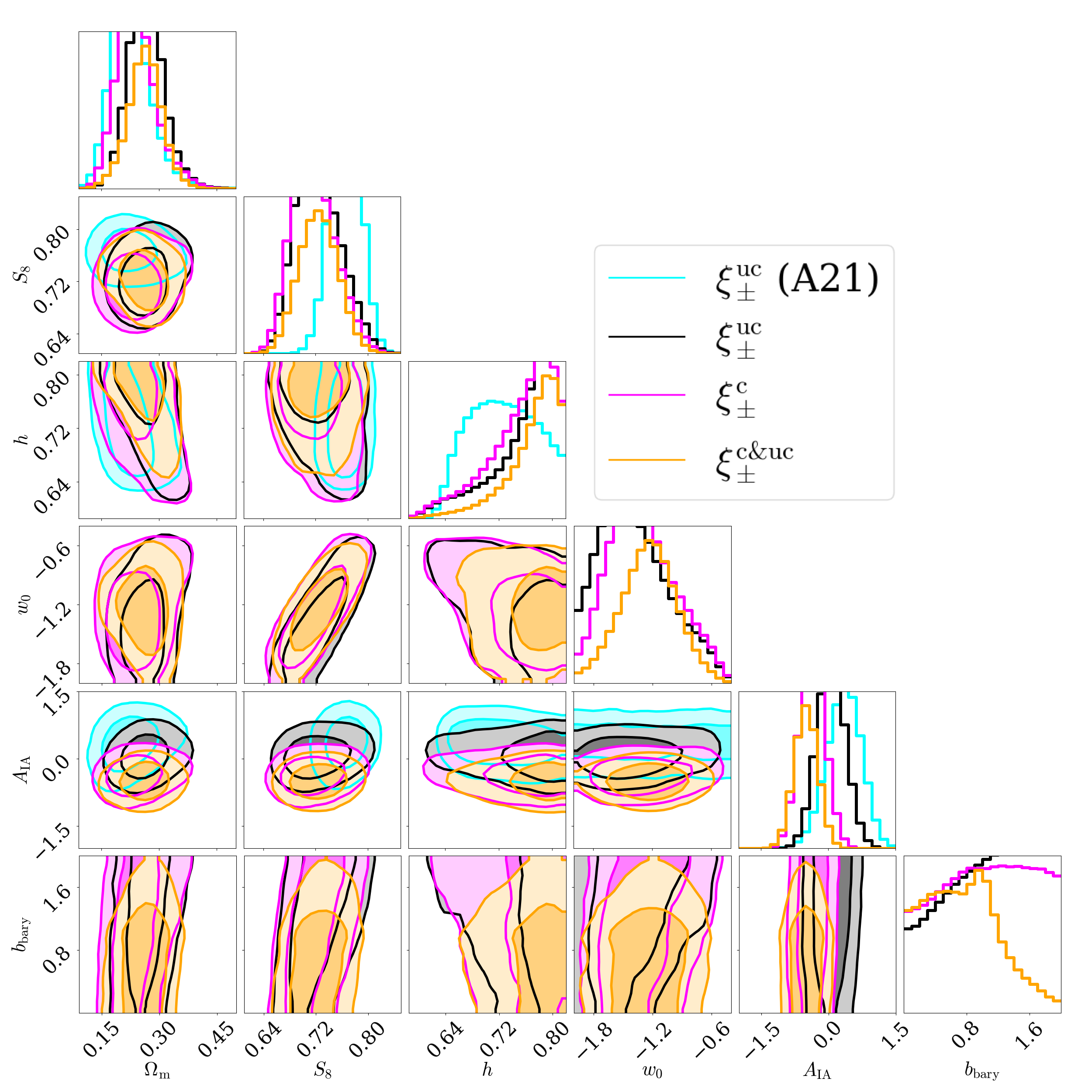}
\caption{Constraints for KiDS-1000 from the clipped (magenta), unclipped (black), and the combined (orange) $\xi_\pm$, compared with the unclipped $\xi_\pm$ constraints of \citet[][A21, cyan]{asgari/etal:2021}. A21 do not constrain $w_0$ (their analysis assumed the $\Lambda$CDM model, $w_0=-1$) or $b_{\rm bary}$; hence there are no cyan constraints for these parameters. }\label{fig:constraints_k1000}
\end{center}
\end{figure*} 

The clipped (magenta), unclipped (black) and combined (orange) constraints for the KiDS-1000 data, including $A_{\rm IA}$ and $b_{\rm bary}$ in addition to the cosmological parameters, are shown in Figure \ref{fig:constraints_k1000}, alongside the unclipped $\xi_\pm$ constraints from A21 (cyan). Our chains were ran with the fiducial inference settings explained in Sec.~\ref{subsec:param_infer}.  We assess the constraining power of the various statistics with the figure of merit, which is inversely proportional to the volume of the 68\% confidence ellipsoid and given by,

\begin{equation} \label{ref:eqn:fom}
{\rm FoM} = \frac{1}{\sqrt{{\rm det}(\boldsymbol{C}_s)}} \,,
\end{equation}

\noindent where $\boldsymbol{C}_s$ is the covariance of the samples obtained by the {\sc nautilus} sampler. When computing the FoM for the entire 11-dimensional parameter space, the 4D cosmological volume, or just the $\Omega_{\rm m}$--$S_8$ plane, we find the combined analysis improves the FoM over the unclipped by factors of 4.0, 1.6 and 1.2 respectively (where the former two are conservative estimates given that the unclipped analysis is prior-limited for $w_0$ but  well constrained in the combined analysis). 

Table \ref{tab:constraints_k1000} presents how these precision gains are distributed across the parameter space: the second column shows the constraint on the four cosmological and seven systematic parameters, with the improvement in precision relative to our unclipped analysis presented in the third column. $\Omega_{\rm m}$ is the only parameter for which the error bar size from the combined analysis is within a percent of that of the unclipped, whereas for all other parameters we observe improvements of 7--32\%. $S_8$ and $w_0$ benefit from 16\% and 24\% gains in precision respectively. Weak lensing is insensitive to the Hubble parameter and so it is unsurprising our analyses returns only a lower limit for $h$; nevertheless, the constraint is still noticeably tighter in the combined analysis than in the unclipped alone.

We find that the nuisance parameters (the full constraints for these, including the $\boldsymbol{\Delta z}$ shifts, are presented in Appendix \ref{sec:appendix_k1000}) are also much more tightly constrained when the traditional approach is augmented with clipping methodology. For example, the intrinsic alignment amplitude constraint, $A_{\rm IA}$, is improved by 27\%. This is most likely due to differences in the way the primarily-large-scale GI and primarily-small-scale II correlations impact the clipped statistic compared with the unclipped. Since clipping suppresses the small-scale contributions to the observed signal, using both the clipped and unclipped measurements helps to break the degeneracy between IA and cosmic shear. This reasoning also applies to the baryonic feedback parameter, $b_{\rm bary}$, which is unconstrained in the unclipped analysis, but the complimentary information from clipping is sufficient to place an upper limit in the combined analysis ($b_{\rm bary}<0.97$). Furthermore, as shown in Appendices \ref{sec:appendix_emu_err} and \ref{sec:appendix_fakedata} for the KiDS-1000 and mock data respectively, when the emulator error is neglected from the covariance (which primarily inflates the errors at small scales), the combined analysis returns a much tighter $b_{\rm bary}$ constraint. These gains in precision likely arises from clipping the high-density peaks, which exist on the same small scales contaminated by BF. This decouples the contributions to the observed signal from different scales. In other words, using the clipped and unclipped probes in tandem allows for more precise determination of scale-dependent processes (both cosmological and astrophysical).

\begin{table}
  \begin{center}
    \caption{The marginalised means and 68\% confidence intervals on the cosmological (upper section) and nuisance (lower section) from the combined (clipped and unclipped) analysis of KiDS-1000. The third column presents the improvements in precision from the combined analysis over the unclipped (the negative figure for $\Omega_{\rm m}$ represents an insignificant broadening of the error bar for this parameter only). The $h$ and $b_{\rm bary}$ constraints are prior-limited and so we do not present improvements for these parameters.} \label{tab:constraints_k1000}
  \begin{tabular}{|lcr|}
\hline
\hline
\textbf{Parameter}  & \textbf{Constraint} & \textbf{(Improvement)} \\ \hline
$\Omega_{\rm m}$ & $0.263^{+0.035}_{-0.038}$ & $-1\%$ \\ \hline
$S_8$ & $0.724^{+0.027}_{-0.027}$ & $16\%$ \\ \hline
$h$ &  $>0.76$ & $-$ \\ \hline
$w_0$ & $-1.24^{+0.26}_{-0.28}$ & $24\%$ \\ \hline \hline
$A_{\rm IA}$ & $-0.50^{+0.19}_{-0.20}$ & $27\%$ \\ \hline
$\Delta z_1$ &  $-0.002^{+0.009}_{-0.009}$ & $ 11\%$ \\ \hline
$\Delta z_2$ &  $0.000^{+0.009}_{-0.009}$ & $ 7\%$ \\ \hline %94\%
$\Delta z_3$ &  $-0.012^{+0.006}_{-0.006}$ & $ 32\%$ \\ \hline %92\%
$\Delta z_4$ &  $-0.008^{+0.007}_{-0.008}$ & $ 7\%$ \\ \hline
$\Delta z_5$ &  $0.008^{+0.006}_{-0.007}$ & $ 28\%$ \\ \hline
$b_{\rm bary}$ & $<0.97$ & $-$ \\ 
\hline \hline
  \end{tabular}
  \end{center}
\end{table}

The comparison of our constraints with the A21 contours  in Fig.~\ref{fig:constraints_k1000} shows an overall consistency across the cosmological parameter space. Our $\Omega_{\rm m}=0.263 ^{+0.035}_{-0.038}$ constraint from the combined analysis of KiDS-1000 is within 0.8$\sigma$ of the A21 constraint\footnote{For ease of comparison, when referencing previous cosmic shear analyses, we always quote the (unclipped) $\xi_\pm$ constraints (specifically, in the case of A21, the ``Best fit + PJ-HPD" approach to identifying the best-fit marginalised value), noting that the fiducial constraints from A21 were derived from COSEBIs (complete orthogonal sets of E/B-integrals).}, $\Omega_{\rm m} = 0.223^{+0.065}_{-0.033}$. The precision of our constraint, 14\%, represents a considerable improvement over A21 (22\%), which we attribute to the improved modelling of non-linear scales in the {\sc cosmoSLICS} simulations relative to the {\sc hmcode}-based modelling of A21.

 \begin{figure} %[t!]
\begin{center}
\includegraphics[width=0.46\textwidth]{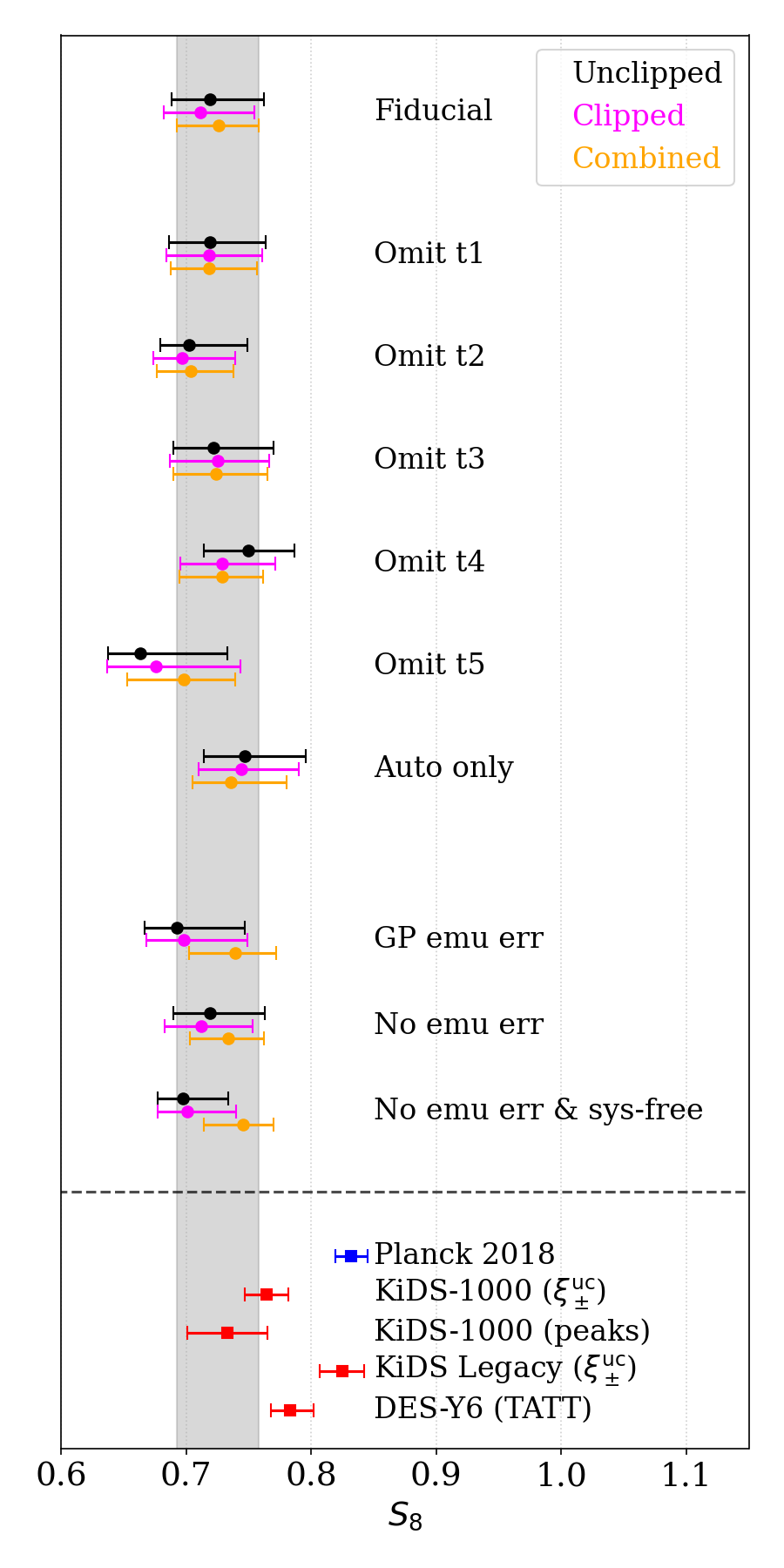}
\caption{$S_8$ constraints for KiDS-1000 under the fiducial inference settings (grey bar) compared with the results from omitting tomographic bins (`t1'--`t5' and `Auto only'), using an alternative definition of the emulator error (`GP emu err'; see Appendix \ref{sec:appendix_emu_err}), or neglecting the error and/or systematics modelling (`No emu err [\& sys-free]') in the inference. The results are compared with $\xi_\pm^{\rm uc}$ constraints from KiDS-1000 (A21), KiDS-Legacy \citep{wright/etal:2025} and DES Year 6 \citep{des/etal:2026}, as well as the peak statistics analysis of KiDS-1000 \citep[][showing only the statistical and not the systematic errors]{harnois-deraps/etal:2024} and the \citet{planck/etal:2018} CMB results.} \label{fig:constraints_k1000_S8}
\end{center}
\end{figure}

Our $S_8$ constraint is in mild agreement with the A21 result at the level of 1.8$\sigma$ and has a precision of 3.7\%. This is competitive with but not outperforming the 2.3\% precision achieved by A21, despite augmenting our analysis with clipping. This is likely due in part to the fact that we have accounted for our modelling (emulator) errors, whereas modelling errors are neglected in the A21 analysis. When we omit the emulator errors from the covariance, our $S_8$ constraint tightens slightly to 3.1\%; this illustrated by the $S_8$ combined (orange) constraints labelled `No emu err' in Figure \ref{fig:constraints_k1000_S8}, which can be compared with the fiducial parameter inference constraints at the top of this figure, represented also by the grey vertical bar. Our analysis of the systematics-contaminated mock data, which neglected the emulator errors, further improved upon this with a 1.9\% constraint on $S_8$ (see Appendix \ref{sec:appendix_fakedata}). The slightly lower precision on $S_8$ relative to A21 may also be due to the fact that we have inferred this parameter within a $w$CDM framework (varying $w_0$) whereas A21 constraints were derived assuming the $\Lambda$CDM model (fixed $w_0=-1$; hence the absence of an A21 constraint on the $w_0$ panels in Fig.~\ref{fig:constraints_k1000}). Our $w_0=-1.24^{+0.26}_{-0.28}$ constraint is consistent with the $\Lambda$CDM model and with a precision of 22\%, slightly weaker than the 12\% precision obtained in the beyond-$\Lambda$CDM analysis of KiDS-1000 by \cite{troester/etal:2020}.

Interestingly, the intrinsic alignment parameter, $A_{\rm IA}=-0.50^{+0.19}_{-0.20}$, with its 27\% gain in constraining power over the unclipped probe alone, has especially high precision compared to A21. This 39\%-precision measurement of the IA amplitude is a significant improvement over the $\sim$90\% precision constraint in A21, $A_{\rm IA} = 0.39^{+0.32}_{-0.37}$, potentially highlighting clipping as an untapped means to better constrain and understand this systematic. We present this argument with caution, however, noting the moderate disagreement between our constraint and that of A21 at the level of $\sim 3.2\sigma$, hinting at the possibility for residual systematics in the KiDS-1000 data, or the necessity for a more complex IA model,  which is not captured by the simulations. On the other hand, our constraints on the shifts in the photometric redshift distributions, presented in Sec.~\ref{sec:appendix_k1000}, are in good agreement with those obtained by A21. The differences in the parameterisation of baryonic feedback between this work and A21, however, means the constraints on the relevant nuisance parameters are not comparable. 

The overall consistency between our constraints and those of A21	 is encouraging, given that our analysis adopts a completely different (simulation- and emulator-based) approach to modelling cosmic shear statistics, their covariance and systematics, and further employs a fully-independent parameter inference pipeline. Obtaining consistent results under such dissimilar frameworks therefore provides an important validation of the methodologies employed in both analyses.

Our best-fit model from the combined analyses under the fiducial parameter inference settings is represented by the black dashed lines in Fig.~\ref{fig:xi}. We measure the goodness-of-fit for this model using the posterior predictive $p$-value evaluated at the best-fit cosmology (the maximum a posteriori, or `MAP', value). This consists of comparing the MAP $\chi^2$ to the distribution obtained for many noise realisations sampled from the covariance. The $p$-value quantifies the probability of obtaining the MAP $\chi^2$ value given the noise distribution of the data, with values below a threshold, taken here to be 0.01, indicating deviations between the model fit and data that cannot reasonably be accounted for by the noise. This method of assessing the $p$-value avoids the need to fit for the effective number of degrees of freedom, which is non-trivial to estimate for large degenerate parameter spaces with informative priors \citep{joachimi/etal:2020}. From the best-fit model in the combined analysis, we obtain a $p$-value of 0.28 for the total (clipped and unclipped $\xi_\pm$) data vector and values of 0.03 and 0.11, respectively, when the calculation is limited to the clipped and unclipped parts of the data vector and covariance. The best-fit models for the clipped and unclipped analyses individually return $p$-values of 0.04 and 0.18 for the respective parts of the data vector. We conclude, therefore, that the clipped, unclipped and combined analyses all return acceptable goodness-of-fit metrics for the KiDS-1000 data.

Figure \ref{fig:constraints_k1000_S8} compares the fiducial $S_8$ constraints (grey bar) with various modifications to the parameter inference settings. From upper to lower, we present the results when various tomographic bins (`t1' to `t5') are omitted from the clipped and unclipped data vectors (including all of their cross correlations with other bins) and when only the auto-correlations are used in the inference. As expected, omitting the highest redshift bins (t4 and t5) has the greatest impact since the galaxies at these redshifts carry the strongest lensing SNR. The shift-up and shift-down when the fourth and fifth bins are omitted, respectively, is also consistent with the findings of A21. The test in which only the auto-correlations were used shows reduced precision relative to the fiducial analyses. This confirms that, despite the signature clipping trough being weaker in the clipped cross-correlations than the auto-correlations (see Fig.~\ref{fig:xi}), the cross-correlations still contribute to the improvement in constraining power from clipping. Encouragingly, there is no sign of internal inconsistency from these tests with all chains returning $S_8$ constraints in statistical agreement.

This is also true for the analysis employing an alternative definition of the emulator error (where the cosmology-dependent Gaussian process covariance is propagated to the likelihood at every step in the inference - see Appendix \ref{sec:appendix_emu_err} for details; labelled `GP emu err'), and in cases where the emulator error is neglected completely (`No emu err' and `No emu err \& sys-free' respectively). The latter results correspond to a case where the IA, BF and photo-$z$ shifts are not modelled in the inference. These reveal the same finding as in A21: accounting for systematics primarily impacts the size of the confidence interval more than the best-fit $S_8$ value itself. Despite the relative insensitivity of the $S_8$ constraints to the emulator error, we find it is necessary to account for the uncertainty in the modelling in order to achieve an acceptable goodness-of-fit.

In the lower part of Fig.~\ref{fig:constraints_k1000_S8}, we compare our results with those of previous analyses: the primary KiDS-1000 (A21), KiDS-Legacy \citep{wright/etal:2025}, and the DES Year 6 \citep[assuming the TATT model for IAs;][]{des/etal:2026} cosmic shear analyses, a peak statistics analysis of KiDS-1000 \citep[based on {\sc cosmoSLICS};][]{harnois-deraps/etal:2024}, and finally the \cite{planck/etal:2018} CMB results. Whilst all of our analysis variations return $S_8$ constraints in agreement with A21, our results invariably prefer lower values of $S_8$, closer to the \cite{harnois-deraps/etal:2024} result. The fact that other HOWLS analyses have also favoured reduced $S_8$ values \citep{sugiyama/etal:2025, gomes/etal:2025} possibly indicates that the additional cosmological information extracted by these probes (and by clipping) genuinely points to  smoother matter clustering in the late-time Universe than is inferred from the \cite{planck/etal:2018} observations.

At face value, Fig.~\ref{fig:constraints_k1000_S8} shows our fiducial combined $S_8$ constraints to be in tension with the Planck result at the level of 3.0$\sigma$. This result should, however, be taken with caution. The comparison of the KiDS-1000 and -Legacy results reveal the impact on $S_8$ from the improved redshift calibration, inclusion of higher redshift sources and wider survey coverage, amongst other improvements implemented in the latter analysis. Hence, it is likely that had the KiDS-Legacy survey specifications been implemented in our simulations, a shift upwards of similar magnitude would have been observed in the unclipped, clipped and combined constraints, thereby reducing the disagreement with Planck.

%% file: Section_5_Conclusions.tex
\section{Conclusions and summary}
\label{sec:conc}

In this analysis, we have measured the first cosmological constraints from applying clipping transformations to weak lensing data, KiDS-1000,  whilst controlling for systematics. Clipping is designed to improve the efficiency of information extraction with two-point statistics by pruning the high-density regions of the observed field and targeting the scales least impacted by baryonic feedback. This work built on the proof-of-concept analysis of \citet[][`G18']{giblin/etal:2018} with numerous methodological improvements.

First of all, the numerical simulations, {\sc cosmoSLICS} we used to model the cosmological dependence of clipped shear correlation functions, $\xi_\pm^{\rm c}$, feature much larger simulation volume, resolution and survey fidelity than were used previously. In particular, the forward modelling of the survey footprint in the simulations mitigated the ``masking bias" seen to affect the $\xi_-^{\rm c}$ in the G18 analysis. This meant that this statistic could be used within the parameter inference alongside the first implementation of tomographic redshift binning of clipped lensing statistics. The clipped data vector was, therefore, greatly expanded relative to the G18 analysis, facilitating greater accessibility of cosmological information. In addition to the survey geometry, we forward modelled the galaxy shape noise, shear calibration, intrinsic alignments, photometric redshift estimation errors and baryonic feedback with suites of dark-matter and hydrodynamical $N$-body simulations tailored to the KiDS-1000 survey specifications. We then trained Gaussian process regression emulators on these simulations to model the cosmological and systematic dependence of the clipped statistics, as well as its conventional counterpart, the unclipped shear correlation function, $\xi_\pm^{\rm uc}$. We demonstrated the accuracy of the emulators to be subdominant to the statistical noise, evaluated with our Covariance Set of simulations, {\sc SLICS}, across almost all angular scales and redshift ranges. Residual emulator error was quantified and factored into the data covariance to avoid potential bias to the inferred cosmological parameters.

We validated the accuracy of our approach using KiDS-1000-like simulated data contaminated with systematics and found the emulators were able to correctly recover the true cosmological and nuisance parameters. We then proceeded to constrain cosmological parameters with the KiDS-1000 data set using both the clipped, unclipped and combined probes. This not only represents the first systematics-controlled clipping analysis of real lensing data, but also a fully independent validation of the KiDS-1000 cosmic shear analysis \citep[][A21]{asgari/etal:2020}, employing emulators trained on numerical simulations opposed to {\sc halofit}-based theoretical modelling. In this analysis we use the same range of angular scales and redshift binning as employed in their (unclipped) $\xi_\pm$ analysis.

We find that combining clipped and unclipped shear correlation functions in the likelihood once again enhances the overall constraining power relative to the traditional approach alone, with the figure of merit in the $\Omega_{\rm m}-S_8$ parameter space improved by a factor of 1.2. The gains in precision of the cosmological constraints, measured relative to the unclipped probe alone,  were strongest for $w_0$ (24\%) and $S_8$ (16\%).The constraints on the Hubble parameter were prior limited in all cases but continued to be tighter in the combined analysis than the unclipped, whilst $\Omega_{\rm m}$ was constrained equally well in both analyses. The systematic parameters were also more tightly constrained when the traditional analysis was augmented with clipping; the intrinsic alignment amplitude was 27\% more tightly constrained in the combined analysis, and an upper limit on the baryonic feedback strength was placed by the combined probe despite it being unconstrained by the unclipped probe alone. 

The emulator-driven cosmological constraints obtained in this work were found to be consistent with the {\sc hmcode}-derived constraints in A21, with higher precision obtained for $\Omega_{\rm m}$  than in A21 (14\% versus 22\%), potentially due to the more accurate modelling of non-linear structures in our simulation training set than is accessible to the {\sc hmcode} predictions. The precision of our $S_8$ constraint is comparable to A21 and our $w_0$ constraint is slightly weaker than the beyond-$\Lambda$CDM analysis performed in \cite{troester/etal:2020} whilst remaining fully consistent with the Standard Model. Our intrinsic alignment constraint is much more precise than the A21 result, but is also in moderate disagreement at the level of $\sim 3.2\sigma$, potentially indicating residual differences between the simulation training set and KiDS-1000 data. As our cosmological constraints - the main focus of this work - remain consistent, however, and our best-fit model returns acceptable posterior predictive $p$-values across the clipped, unclipped and combined analyses, we leave further investigation of the impact of clipping on intrinsic alignments for future work.  

This work has progressed the use of clipping in cosmic shear from the realm of `proof-of-concept' forecasts to a competitive and robust lensing statistic, applicable to concurrent data sets. Challenges remain to unlock its full potential, however. The residual inaccuracies in emulator predictions, arising from the expense of large simulation suites sampling the cosmological and systematic parameter space, particularly impacted the small-scale measurement and hence, the constraints on the baryonic feedback strength. This obstacle could be overcome in future with suites such as {\sc Gower St} \citep{jeffrey/etal:2025} and {\sc CosmoGrid} \citep{kacprzak/etal:2023}, which sample more cosmological parameters and with higher node density (albeit at the cost of spatial resolution in the latter case).

Additionally, we adopted a fixed clipping threshold in this analysis, motivated by the tests in G18. Future analyses could look to optimise this free parameter and extract even more cosmological information from the data. The upcoming release of Stage-IV lensing data from the Euclid Space Telescope and later, the Vera Rubin Legacy Survey of Space and Time (LSST), also present excellent opportunities to further leverage the enhancements to cosmic shear constraints from clipping.

The significant gains in precision we have seen for the intrinsic alignment and baryonic feedback parameters, made possible by the decoupling of scale-dependent information through clipping, highlight the potential for this transform to be used in better understanding astrophysical processes relevant to cosmology analyses. For example, clipped cosmic shear constraints may help to refine hydrodynamical simulations, acting as a complimentary probe to $X$-ray cross-correlations \citep{ferreira/etal:2024} in shedding light on the uncertainty surrounding BF and its impact on cosmological statistics \citep{chisari/etal:2018}. Future work in this direction could also bring together clipped lensing and clustering \citep{simpson/etal:2011, simpson/etal:2013, wilson:2016} to create a ``clipped $3\times 2$-point" analysis which fully capitalises on the wealth of constraining power in these probes. 

Clipping remains a fairly untapped source of cosmological and astrophysical information. With the upcoming advancements in data quality and simulation resources, the remaining obstacles to fully utilising this approach are likely to be short-lived.

%% file: Section_Appendix.tex
\section{Incorporating Emulator Error} \label{sec:appendix_emu_err}

 \begin{figure*}
\begin{center}
\includegraphics[width=0.8\textwidth]{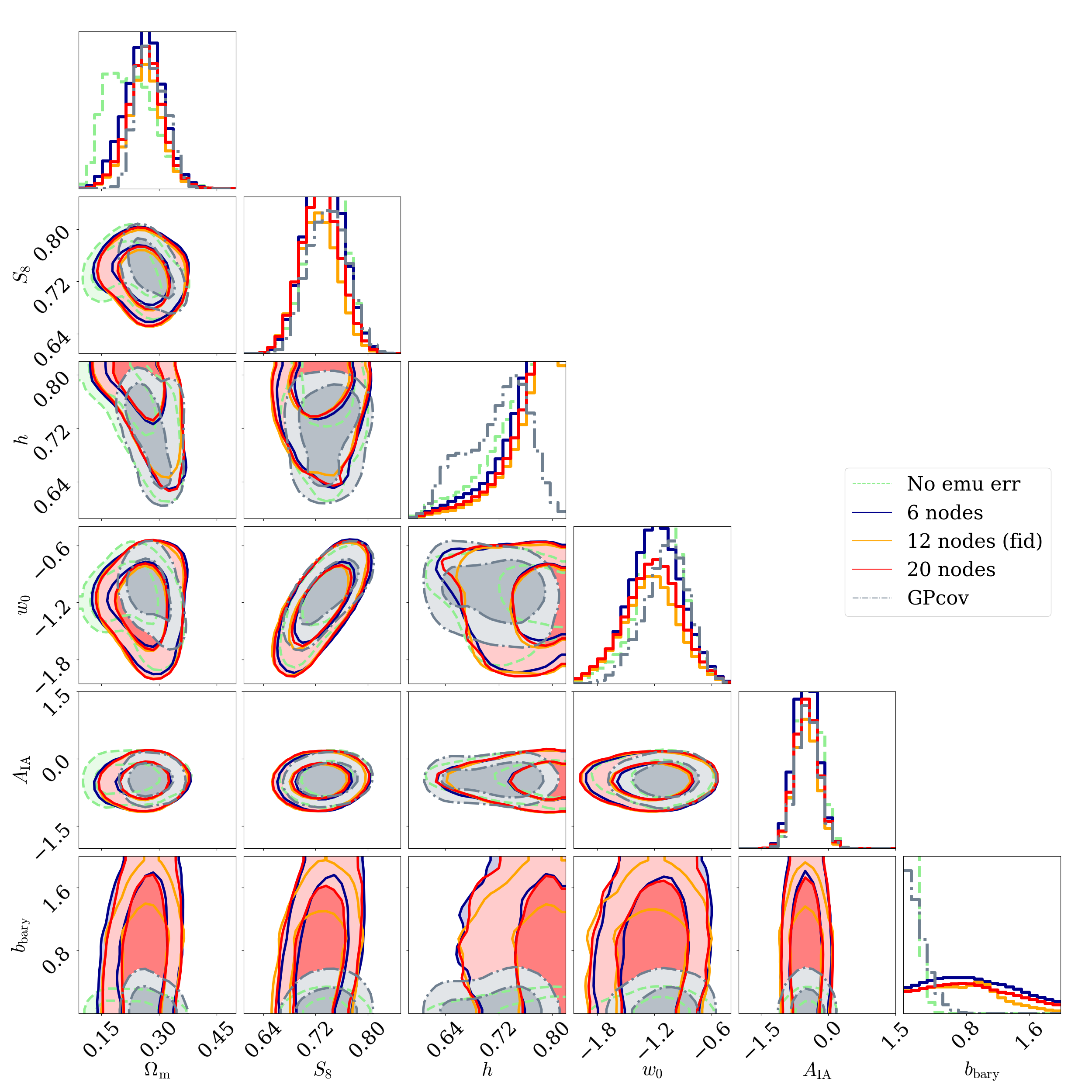}
\caption{Constraints on the cosmological parameters and the strengths of IA and BF from the combined analysis of KiDS-1000 for different definitions of the emulation error: no emulator error (dashed green), increasing the number of central cosmology nodes used to evaluate the emulator uncertainty (blue, orange, red), or propagating the cosmology-dependent GP covariance to the likelihood at each step in the inference (dot-dashed grey).}\label{fig:emu_errs}
\end{center}
\end{figure*} 

The optimal method for incorporating emulator errors in parameter inference is an open question. Our fiducial approach averages the squared differences between the emulated and simulation $\xi_\pm^{\rm c/uc}$ obtained in leave-one-out cross-validation (CV) and augments the diagonal of the covariance matrix (Eqn.~\ref{eqn:Cov_emu}). We experimented with the number of nodes, $n$, in the Cosmology Set used in this averaging, conscious of the fact that inclusion of the outermost nodes (where emulators are tasked with extrapolating beyond the training set) presents a pessimistic appraisal of the emulator accuracy, but also aiming to account for emulation accuracies in all of the regions of interest within the parameter space. We found that our constraints are insensitive to the number of nodes used to define the emulator error. Specifically, we tried the following scenarios:

\begin{itemize}

\item $\boldsymbol{n=20}$: All nodes except for those which have a maximum or minimum value of $\Omega_{\rm m}$, $S_8$, $h$, or $w_0$ were used to average the emulator inaccuracy. This definition, in practice, excludes 6 of the 26 nodes in the cosmology training set and is equivalent to the approach of \cite{harnois-deraps/etal:2024}. 

\item $\boldsymbol{n=12}$ (the fiducial analysis): This averages the emulator errors of the 12 nodes closest to the fiducial cosmology (including the fiducial itself, shown by the black star in Fig.~\ref{fig:nodes}). This was defined in terms of the Euclidean distance between the nodes and the fiducial in a dimensionless unitary scaling of the parameter space (i.e. with all cosmological values scaled by the range of the training set to be $\in [0,1]$).

\item  $\boldsymbol{n=6}$: the same as above but using only the 6 closest nodes to the fiducial cosmology.

\item $\boldsymbol{n=0}$: equivalent to neglecting the emulator error.

\end{itemize}

Figure \ref{fig:emu_errs} shows the cosmological constraints alongside those on the IA and BF strengths from the combined analysis using these four definitions of the emulator error. We find that all values of $n>0$ ($n=6$, dark blue; the fiducial $n=12$, orange; $n=20$, red) return very similar constraints, revealing that the best-fit cosmology and precision of the constraining power is robust to how many nodes are used to define the parameter space over which the emulator error is evaluated. The cosmological constraints where the emulator error is neglected ($n=0$, dashed green) are also consistent with these other constraints, but have slightly worse precision for some parameters (e.g. $\Omega_{\rm m}$). This is perhaps surprising, given these constraints were obtained using a covariance with diagonal elements which are a few \% smaller than those used in the $n>0$ analyses. We attribute this broadening to noise in the sampling of the likelihood surface, whereas the inflated covariance matrices ($n>0$) lead to smoother likelihoods with less spurious noise features. 

Interestingly, when the emulator error is neglected, the precision of the $b_{\rm bary}$ constraint improves dramatically. This is because the emulator error for the clipped statistic is larger on small scales (where BF dominates) due to the challenge of learning the non-trivial shape of the clipping trough. Including the emulator error therefore dilutes the complimentary information on BF provided by the clipped probe. We do claim the $b_{\rm bary}$ constraint obtained in the absence of accounting for the emulator error is necessarily unbiased (this analysis returns a poor goodness-of-fit), but it highlights the exciting potential for constraining baryonic physics with clipping in future simulation suites featuring lower emulation uncertainty.

These approaches to evaluating the emulator error change only the diagonal of the covariance (or not at all in the case of $n=0$) when in principle, emulator errors could be correlated across angular scales\footnote{The use of per-statistic and per-tomographic-bin emulators, however, negates the possibility of correlations in emulator errors across these parts of the prediction vector.}. One could, in principle, approximate the cross-covariance of the emulator errors between $\theta$-bins with the CV predictions, but the number of predictions (equal to the number of nodes, 26) is relatively small compared to the data vector lengths predicted by each emulator ($n_\theta = 9$), leading to a noisy estimate of the emulator error covariance. The approaches outlined above also assume an emulator error averaged across the parameter space is representative of the error at each step in the parameter inference, when in fact the accuracy varies with position (it decreases with increasing distance from the training nodes). To evaluate whether the cross-covariance or cosmology dependence of the emulator errors could impact cosmological constraints, we investigated propagating the Gaussian process (GP) covariance at every step in the inference to the likelihood. In practice, this meant that the covariance appearing in our Gaussian likelihood (Eqn.~\ref{eqn:Lhd_Gauss}) varies as the {\sc nautilus} chains explored the parameter space.

The GP covariance is given by
\begin{equation}
\mathbf{C}^{\rm GP}(\mathbf{f}_*) = k(\mathbf{X}_*, \mathbf{X}_*) - k(\mathbf{X}_*, \mathbf{X}) \left[ k(\mathbf{X}, \mathbf{X}) \right]^{-1} k(\mathbf{X}, \mathbf{X}_*) \,, 
\end{equation}
where $\mathbf{f}_*$ is the GP prediction for a coordinate $\mathbf{X}_*$ in the parameter space, $\mathbf{X}$ is the array of training nodes (for the cosmological emulators, this has dimensionality $N \times d$, for the $N=26$ distinct cosmologies in {\sc cosmoSLICS} spanning the $d=4$ $\left[ \Omega_{\rm m}, S_8, h, w_0 \right]$ volume). $k(\mathbf{X}, \mathbf{X})$ is the $N \times N$ matrix describing the covariance between training nodes,
\begin{equation}
k(\mathbf{X}, \mathbf{X}) = \begin{bmatrix} 
k(\mathbf{x}_1, \mathbf{x}_1) & k(\mathbf{x}_1, \mathbf{x}_2) & \cdots & k(\mathbf{x}_1, \mathbf{x}_N) \\ 
k(\mathbf{x}_2, \mathbf{x}_1) & k(\mathbf{x}_2, \mathbf{x}_2) & \cdots & k(\mathbf{x}_2, \mathbf{x}_N) \\ 
\vdots & \vdots & \ddots & \vdots \\ 
k(\mathbf{x}_N, \mathbf{x}_1) & k(\mathbf{x}_N, \mathbf{x}_2) & \cdots & k(\mathbf{x}_N, \mathbf{x}_N) 
\end{bmatrix} \,,
\end{equation}
where the kernel used in this application is the radial basis function (Gaussian),
\begin{equation}
k(\mathbf{x}_i, \mathbf{x}_j) = \sigma_f^2 \exp\left( -\frac{1}{2} \sum_{m=1}^{d} \frac{(x_{i,m} - x_{j,m})^2}{\ell_m^2} \right) .
\end{equation}
Here, $\sigma_f^2$ and $\ell_m$ are the hyperparameters: the kernel amplitude and the correlation lengths per dimension, respectively \citep{wang:2020}. 

Adding the covariance of the GP prediction itself at each step in the chain, $\mathbf{C}^{\rm GP}$, to the covariance measured from {\sc SLICS} accounts for the variation in the emulation error as a function of position in the parameter space, and the covariant errors across the $\theta$ bins of the prediction. We do not, however, account for the errors from the systematics emulators as they are subdominant to those of the cosmological emulators. The use of separate cosmological emulators for each redshift bin and statistic means that the {\sc SLICS} covariance is altered by the GP covariance within $n_\theta \times n_\theta$ redshift-bin blocks\footnote{One further deviation between this analysis and the others performed in this work, is that no PCA was performed on the cosmology training set as this greatly simplifies propagation of the GP covariance. We tested that removing the PCA had no impact on the cosmological and systematic constraints, and is hence, not important for maintaining emulation accuracy}.

The constraints corresponding to this alternative definition of the emulation error are shown by the dot-dashed grey contours in Fig.~\ref{fig:emu_errs}. These are consistent with and more highly constraining than the contours from the other emulator error definitions, suggesting that accounting for the cosmology-varying GP covariance is a more efficient method for handling the modelling error from emulators. Moreover, this approach returns very tight $b_{\rm bary}$ constraints, similar to the case where the emulator errors were neglected. This approach is not without its disadvantages, however, as it required the re-inversion of the $\boldsymbol{C}+\boldsymbol{C}^{\rm GP}$  matrix at each step in the inference, which increased the sampling time by a factor of 3.5. This meant that it would be prohibitively expensive to adopt this emulator error definition for all chains ran in this work. Furthermore, whilst the combined constraints shown here appear stable, the individual clipped and unclipped constraints obtained from propagating the GP covariance showed noticeably noisier likelihood surfaces than those assuming other definitions for the emulator error. This could be due to instabilities in the covariance inversion emerging as the covariance is varied during parameter inference.

For these reasons, we chose to adopt the more stable and time-efficient definition of the emulator error in our fiducial analysis, but note that our primary constraints presented in the main body of this paper (especially for $b_{\rm bary}$) can be regarded as conservative, given more efficient handling of the GP covariance is possible in principle, as evidenced by the grey contours in Fig.~\ref{fig:emu_errs}. This approach is likely to be more feasible in future analyses given larger mock suites which will facilitate more stable computations of both the simulated and GP covariance matrices.

\section{Impact of Source-Lens Clustering} \label{sec:appendix_slc}

As discussed in Sec.~\ref{subsec:sims}, the positions of galaxies in the IA Set of simulations trace the underlying dark matter density with a redshift-independent linear galaxy bias of $b_{\rm gal}=1.0$. These mocks allow us to test the impact of source-lens  clustering (SLC) on our cosmological constraints. SLC occurs preferentially along more overdense lines of sight, at which the lensing signal is also expected to be higher, leading to a biased sampling of the shear field. SLC means the lensing signal becomes correlated with both the effective galaxy number density, and with the intrinsic galaxy shape noise (due to its connection to the local underlying density). \cite{krause/etal:2021} found this higher-order effect is negligible on cosmic shear constraints derived from Stage-III surveys, and hence is not expected to impact those derived from our unclipped correlation functions (we verify this conjecture in this section). 
\cite{gatti/etal:2024}, however, found significant detections of SLC in some HOWLS measured from DES Year 3 data. This motivates this investigation of whether SLC affects clipped correlation functions, which have some similarities with HOWLS in terms of their capacity to extract additional cosmological information through non-linear transformations of the observed field.

SLC modulates galaxy shears, $\gamma$, and intrinsic ellipticities, $\epsilon_{\rm int}$, at galaxy angular positions, $\boldsymbol{\theta}_g$, as follows:
\begin{equation}
\begin{gathered}
   \gamma^{\rm SLC}\left( \boldsymbol{\theta}_g \right) = \left[ 1+ b_{\rm gal} \kappa(\boldsymbol{\theta}_g) \right] \gamma\left( \boldsymbol{\theta}_g \right) \,, \\
    \epsilon_{\rm int}^{\rm SLC}\left( \boldsymbol{\theta}_g \right) = \frac{\epsilon_{\rm int} \left( \boldsymbol{\theta}_g \right)}{ \sqrt{1+ b_{\rm gal} \kappa(\boldsymbol{\theta}_g) }  } \,,
\end{gathered}
\end{equation}
where we present the prescription from \cite{gatti/etal:2024} integrated over redshift. The strength of SLC is therefore controlled by the level of galaxy bias. Given that the shape noise is added to our simulated galaxies in post-processing, it is straight forward to modulate the impact of SLC on the simulated intrinsic galaxy ellipticities simply by toggling the value of $b_{\rm gal}$. The shear values we measure from the IA simulations, however, are already contaminated by SLC with a hard-coded value of $b_{\rm gal}=1.0$; hence, we rescale them accordingly:
\begin{equation}
 \widetilde{\gamma^{\rm SLC}}\left( \boldsymbol{\theta}_g \right) = \left( \frac{1+b_{\rm gal}\kappa(\boldsymbol{\theta}_g)}{1+\kappa(\boldsymbol{\theta}_g)} \right) \gamma^{\rm SLC}\left( \boldsymbol{\theta}_g \right) .
\end{equation}

\begin{figure}
\begin{center}
\includegraphics[width=0.5\textwidth]{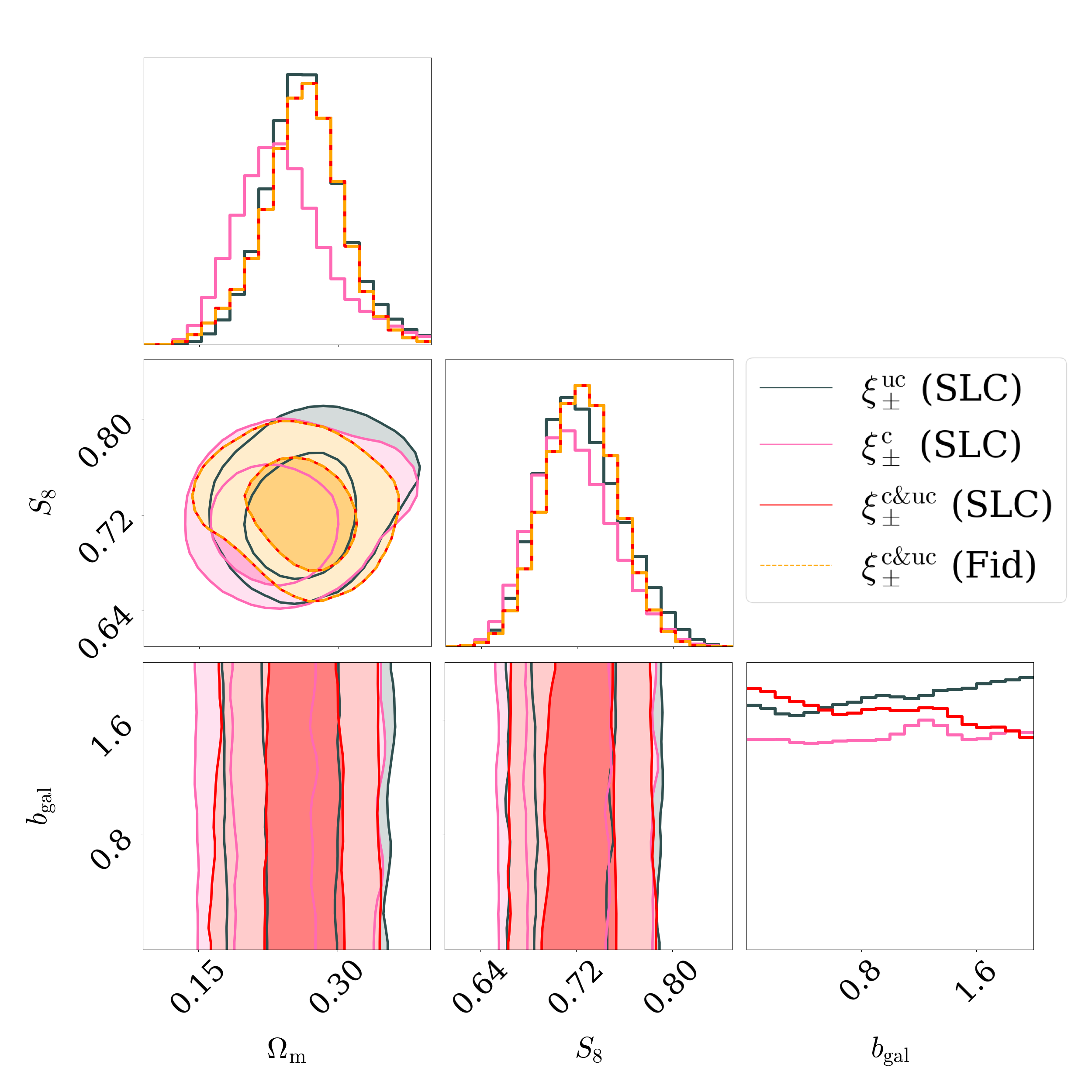}
\caption{The galaxy bias, $b_{\rm gal}$, remains unconstrained in the unclipped/clipped/combined (grey/pink/red) analyses when included as a nuisance parameter, whilst the constraints on the other parameters remain unchanged (here showing only $\left[ \Omega_{\rm m}, S_8 \right]$) relative to the fiducial analysis (here showing the fiducial combined constraint in dashed orange) which assumes a fixed $b_{\rm gal}=1.0$.}\label{fig:slc_constraints}
\end{center}
\end{figure}

We compute the bias to the clipped and unclipped $\xi_\pm$ from SLC in the same way as the other systematics discussed in Sec.~\ref{subsec:mod_sys},
\begin{equation}
B_{\rm SLC} = \xi_{b_{\rm gal}} - \xi_{(b_{\rm gal}=0)} \,,
\end{equation}
for $b_{\rm gal}=[0.75, 1.0, 1.25]$, finding biases to the clipped and unclipped $\xi_\pm$ (relative to the benchmark, $b_{\rm gal}=0$) at the level of $\lesssim 0.5\%$, with very little dependence on $ b_{\rm gal}$. This implies both statistics are highly robust to the presence of SLC, but we test this further by fitting a separate linear model to $B_{\rm SLC}$ for each $\theta$ bin, redshift bin and statistic, in the same manner as we did for the baryonic feedback, and including $b_{\rm gal}$ as an additional nuisance parameter in the sampling. We find it to be unconstrained and with the cosmological/nuisance parameter constraints unchanged by its inclusion. This is evidenced by Figure \ref{fig:slc_constraints} which compares the unclipped (grey), clipped (pink) and combined (red) constraints on $\Omega_{\rm m}$, $S_8$ and $b_{\rm gal}$ from this test with the combined constraints from the fiducial parameter inference settings (dashed orange contour) outlined in Sec.~\ref{sec:results} which did not vary $b_{\rm gal}$. The combined constraints are practically identical irrespective of the inclusion of $b_{\rm gal}$ as a nuisance parameter.

\section{Additional constraints} 

\subsection{Mock data constraints} \label{sec:appendix_fakedata}

 \begin{figure*}
\begin{center}
\includegraphics[width=\textwidth]{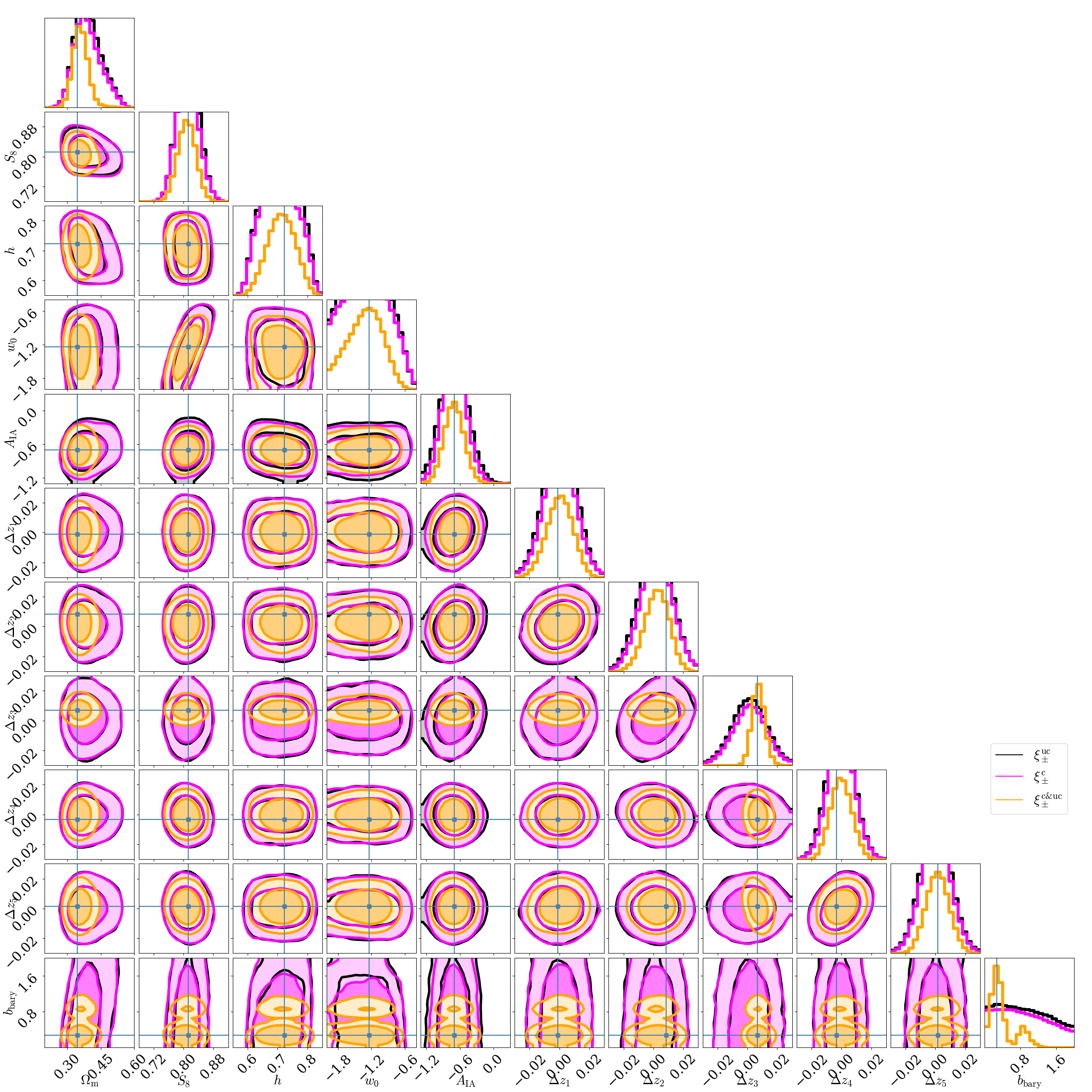}
\caption{Constraints for the systematics-contaminated mock data (with input parameters shown by the blue cross-hair) from the unclipped (black), clipped (magenta) and combined $\xi_\pm$ (orange).}\label{fig:constraints_trial42}
\end{center}
\end{figure*}

Using our fiducial parameter inference settings, we validate the accuracy of our modelling by obtaining constraints for two synthetic data sets with known cosmologies. One of these is the systematics-free average $\xi_\pm^{\rm c/uc}$ from the Covariance Set, the constraints from which verify whether our modelling recovers the true cosmology for data which is not present in the training set, and the other is a systematics-contaminated data vector generated by our trained emulators and linear models with cosmological and nuisance parameters sampled from the priors. As the input parameters do not match any of those used in the simulations, this second data is used to test the accuracy of the interpolation between nodes and verify that cosmological and systematic contributions to the $\xi_\pm^{\rm c/uc}$ can be successfully disentangled in the parameter inference.

When running chains on the synthetic data, we neglect the emulator error (discussed in Sec.~\ref{subsec:emu_train}) from the covariance. Recovering the true input parameters under these conditions presents a more difficult test for the modelling to pass and also demonstrates the achievable improvement in constraining power from clipping given future simulation suites with negligible emulation error. Our fiducial parameter inference settings, however, applied to the KiDS-1000 data, do use the covariance modulated by the emulation error (via Eqn.~\ref{eqn:Cov_emu}).  

\begin{table}
  \begin{center}
    \caption{The marginalised means and 68\% confidence intervals (centre-right column) on the cosmological (upper section) and nuisance (lower section) parameters from the combined analysis of the systematics-contaminated synthetic data. All input values (centre-left column) are successfully recovered and the gains in precision over the unclipped-alone are presented in the final column.  } \label{tab:constraints_trial42}
  \begin{tabular}{|lccr|}
\hline
\hline
\textbf{Parameter} & \textbf{Input}  & \textbf{Constraint} & \textbf{Improvement} \\ \hline
$\Omega_{\rm m}$ & 0.346 & $0.357^{+0.027}_{-0.024}$ & $54\%$ \\ \hline
$S_8$ & 0.813 & $0.808^{+0.021}_{-0.021}$ & $ 18\%$ \\ \hline
$h$ & 0.724 &  $0.713^{+0.045}_{-0.046}$ & $ 28\%$ \\ \hline
$w_0$ & -1.24 & $-1.30^{+0.28}_{-0.35}$ & $ 24\%$ \\ \hline \hline
$A_{\rm IA}$ & -0.70 & $-0.71^{+0.16}_{-0.16}$ & $ 30\%$ \\ \hline
$\Delta z_1$ & -0.001 & $0.000^{+0.008}_{-0.008}$ & $ 19\%$ \\ \hline
$\Delta z_2$ & 0.008 & $0.002^{+0.007}_{-0.008}$ & $ 26\%$ \\ \hline %94\%
$\Delta z_3$ & 0.007 & $0.007^{+0.003}_{-0.003}$ & $ 73\%$ \\ \hline %92\%
$\Delta z_4$ & -0.003 & $0.000^{+0.007}_{-0.006}$ & $ 20\%$ \\ \hline
$\Delta z_5$ & 0.002 & $0.001^{+0.007}_{-0.007}$ & $ 22\%$ \\ \hline
$b_{\rm bary}$ & 0.38 & $0.29^{+0.55}_{-0.08}$ & $ >85\%$ \\ 
\hline \hline
  \end{tabular}
  \end{center}
\end{table}

The cosmological and systematic constraints from the clipped (magenta contour), unclipped (black) and combined (orange) analyses of the systematics-contaminated synthetic data are presented in Figure \ref{fig:constraints_trial42}. We find that the modelling successfully recovers the true values in all cases even when the emulator errors (see Sec.~\ref{subsec:emu_train}) are neglected, and that the combined $\xi_\pm^{\rm c/uc}$ offers significantly improved precision over the unclipped $\xi_\pm$ alone: the $[\Omega_{\rm m}, S_8, h, w_0]$ 68\%-confidence constraints are [54\%, 18\%, 28\%, 24\%] tighter respectively. The constraints are presented numerically, along with their improvements over the $\xi_\pm^{\rm uc}$ alone, in Table \ref{tab:constraints_trial42}.

The enhancement in constraining power facilitated by clipping in this test is on average higher than the $\sim 17\%$ gains measured for $S_8$ in G18. We attribute this difference to the many improvements made in the clipping methodology in this work, including improved simulations, implementing tomography and using $\xi_-^{\rm c}$ in addition to $\xi_+^{\rm c}$ for the first time in parameter inference. We also find that the combined $\Omega_{\rm m}$ constraint is considerably tighter than the unclipped, in contrast to the corresponding results from the KiDS-1000 data, for which the constraints were of equal precision. This may be in part due to the inclusion of the emulator error in the analysis of KiDS-1000 (neglected here), which on average is higher for the clipped parts of the emulated data vector than the unclipped (thereby affecting the clipped and combined constraints more than the unclipped alone). It may also imply that the details of the precision gains, on a parameter-by-parameter basis, are to some extent influenced by the specifics of the data and the input cosmology's proximity to simulated nodes. The overall picture, however, is clear: the combined analysis facilitates tens of percent improvements in precision over the unclipped alone.

This is evidenced also by the improvements in the figures of merit (FoM; Eqn.~\ref{ref:eqn:fom}) from the combined analysis over the unclipped: computed for the $\Omega_{\rm m}$--$S_8$ plane, 4D cosmological, and 11D total parameter volumes, the FoMs are improved by factors of 2.3, 3.5, and 90, illustrating the potential power of clipping for future analyses with negligible emulator errors. Contributing largely to the improvements in FoM are the gains in precision for systematic parameters, particularly $b_{\rm bary}$ (>85\%\footnote{As the unclipped 68\% confidence interval on $b_{\rm bary}$ is larger than the prior bounds, the gain in precision here is computed by comparing the combined constraint to the full prior range (and hence represents a conservative estimate of the improvement).}), which is unconstrained in the individual clipped and unclipped analyses but is constrained by their combination, which we attribute to the decoupling of scale-dependent information (discussed more in Sec.~\ref{sec:results}) in the analysis benefitting from access to the information contained in both statistics. A second, lesser peak is seen in the marginalised posterior on this parameter from the combined probe, somewhat broadening the upper bound. This is caused by noise in the likelihood surface which is largely mitigated by the inclusion of emulator error in the fiducial analysis. Unfortunately, including the emulator error also greatly dilutes the $b_{\rm bary}$ constraining power overall, as shown in Appendix \ref{sec:appendix_emu_err}. Our results with the systematics-contaminated mock data confirm those obtained for KiDS-1000: that potentially high-precision baryonic feedback constraints are achievable from future clipped lensing analyses utilising improved emulators.

\begin{figure}
\begin{center}
\includegraphics[width=0.5\textwidth]{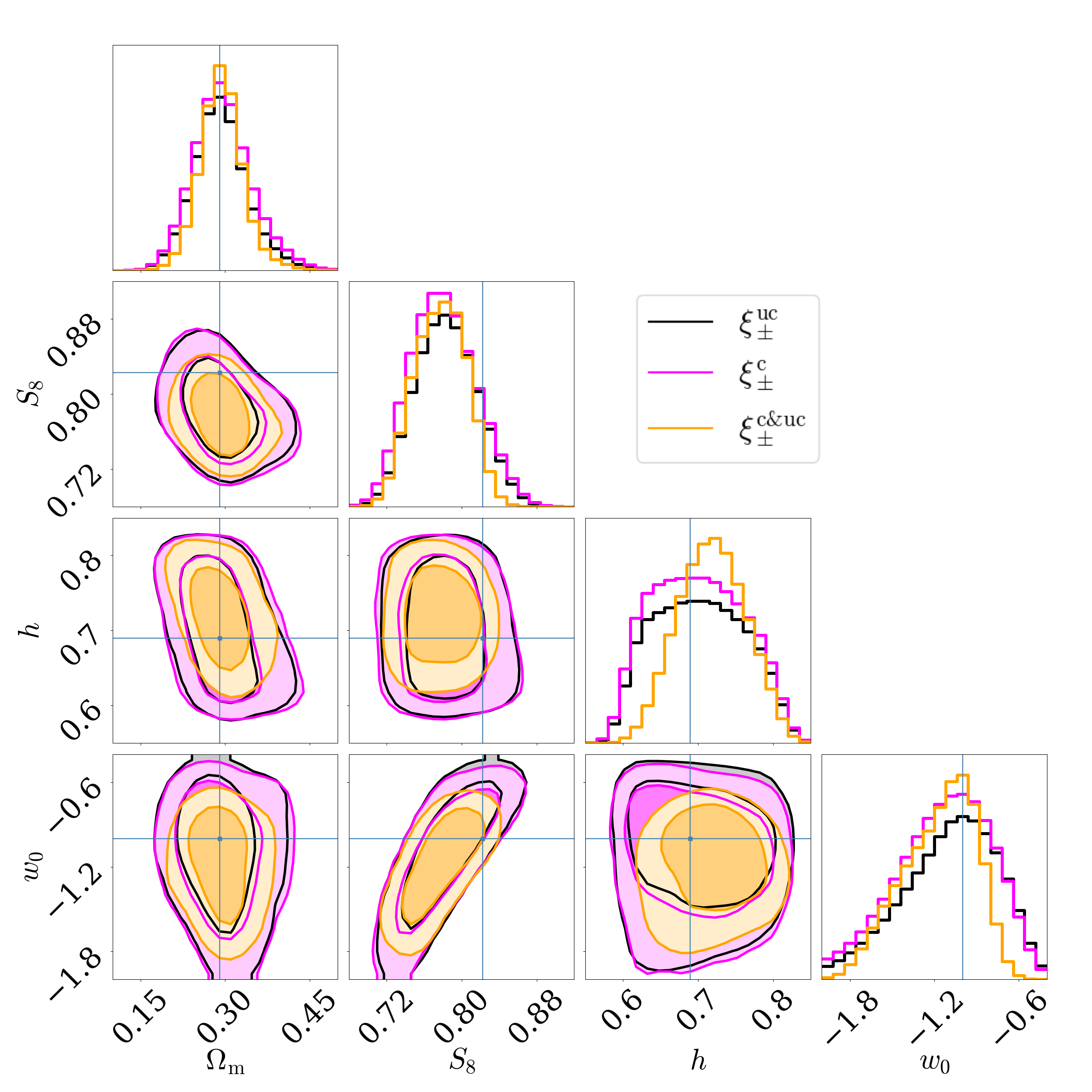}
\caption{Cosmological constraints for the systematics-free synthetic data (SLICS) with the true cosmology shown by the blue cross-hair. This mock did not serve as part of the training set and hence tests the capacity of our cosmological emulators to generalise to external simulations. }\label{fig:constraints_slics}
\end{center}
\end{figure}

Figure \ref{fig:constraints_slics} presents the cosmological constraints for our alternative mock data: the average of the Covariance Set simulations ({\sc SLICS}). As these measurements are not part of the Cosmology Set, nor derived from them (as is the case for the systematics-contaminated mock data), this data can be used to verify that unbiased constraints are obtained even when the emulators are tasked with generalising beyond their training data (as they are required to do for KiDS-1000). Although, being systematics free, this data can, by definition, only verify this for the cosmological parameters. 

We again perform these tests using the fiducial parameter inference settings (see Sec.~\ref{subsec:param_infer}), but neglecting the emulator error as we did with the systematics-contaminated mock data. This presents a more stringent and difficult test of the emulators' generalisability. We find that the true values (blue cross-hair) continue to be recovered to within $\simeq$1$\sigma$ in all analyses: unclipped (black), clipped (magenta) and their combination (orange). The improvements from the combined analysis relative to the unclipped are slightly smaller than those observed for the cosmological parameters of the systematics-contaminated mock data (Tab.~\ref{tab:constraints_trial42}): $[-19,18,12,15]\%$ for $[\Omega_{\rm m}, S_8, h, w_0]$ respectively (with a 1.4 factor improvement in the FoM in the $\Omega_{\rm m}$--$S_8$ plane). This confirms what was observed for the systematics-contaminated mock and KiDS-1000 data vectors - that the $\Omega_{\rm m}$ constraint is most sensitive to the specific shape of the input data vector. This test also serves its main purpose, which is to validate the generalisability of our modelling beyond the training set simulations.

\subsection{KiDS-1000 systematic constraints} \label{sec:appendix_k1000}

\begin{figure*}
\begin{center}
\includegraphics[width=0.80\textwidth]{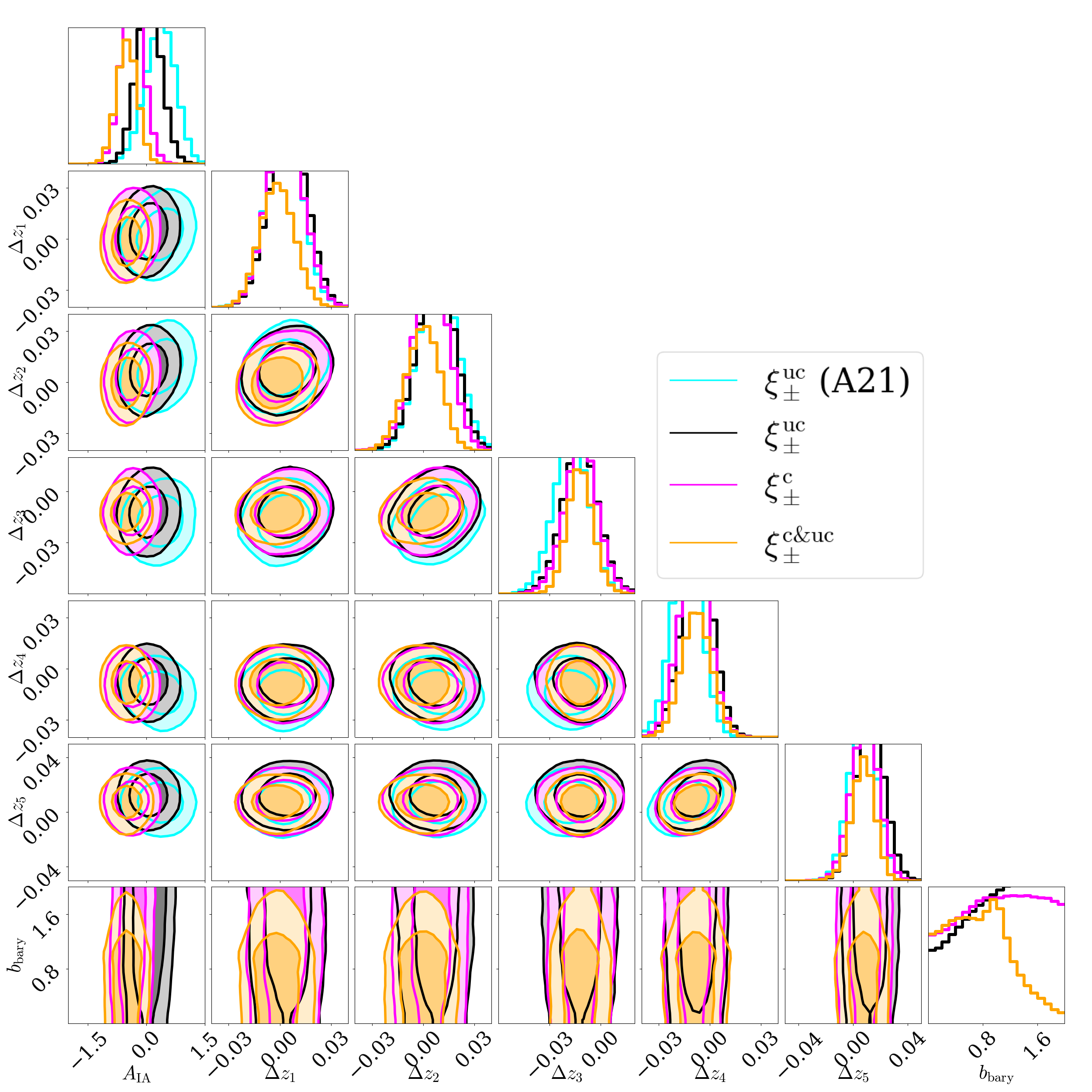}
\caption{Systematic constraints for KiDS-1000 from the clipped (magenta), unclipped (black), and the combined (orange) $\xi_\pm$, compared with the unclipped $\xi_\pm$ constraints of \citet[][A21, cyan]{asgari/etal:2021}. Since A21 use a different parameterisation of baryonic feedback, no A21 constraint is presented for $b_{\rm bary}$. The corresponding cosmological constraints are shown in Fig.~\ref{fig:constraints_k1000}.}\label{fig:constraints_k1000_sys}
\end{center}
\end{figure*}

The constraints on the nuisance parameters from the unclipped (black), clipped (magenta) and combined (orange) analyses of KiDS-1000 are shown in Figure \ref{fig:constraints_k1000_sys}, complimenting the cosmological constraints presented in Fig.~\ref{fig:constraints_k1000}. As discussed in Sec.~\ref{sec:results}, consistent constraints with those of A21 (cyan) are seen for the five mean shifts in the photometric redshift distributions, $\Delta z_i$. Our baryonic feedback parameter, $b_{\rm bary}$, is not comparable with the corresponding parameter used to marginalise over this effect in A21 \citep[the {\sc hmcode} parameter, $A_{\rm bary}$, of][]{mead/etal:2016}. Hence, no A21 constraint is shown for $b_{\rm bary}$ here. Finally, the intrinsic alignment amplitude, $A_{\rm IA}$, constraints present the greatest difference between our combined and the A21 analyses (3.2$\sigma$). As discussed in Sec.~\ref{sec:results}, differences in marginalised posteriors are expected to some extent, especially given our modelling pipelines are completely independent. We are also confident, given our tests with the systematics-free and -contaminated mock data (Appendix \ref{sec:appendix_fakedata}), that our emulators obtain robust cosmological constraints (the main focus of this work). As such, we leave further investigation of the impact of any differences in the IA modelling between A21 and our approach to future work.